\documentclass[iop]{emulateapj}
\usepackage{amsmath,amssymb,amstext,ulem}
\usepackage{aas_macros}
\usepackage{natbib}
\shorttitle{Searching for Wide Companions}
\shortauthors{Martinez \& Kraus}

\begin{document}

\title{Searching for Wide Companions and Identifying Circum(sub)stellar Disks through PSF-Fitting of \textit{Spitzer}/IRAC Archival Images}

\author{Raquel A.\ Martinez, Adam L.\ Kraus}
\affil{Department of Astronomy, The University of Texas at Austin, Austin, TX 78712, USA}

\begin{abstract}
Direct imaging surveys have discovered wide-orbit planetary-mass companions that challenge existing models of both star and planet formation, but their demographics remain poorly sampled. We have developed an automated binary companion point spread function (PSF) fitting pipeline to take advantage of \textit{Spitzer}'s infrared sensitivity to planetary-mass objects and circum(sub)stellar disks, measuring photometry across the four IRAC channels of 3.6 $\mu$m, 4.5 $\mu$m, 5.8 $\mu$m, and 8.0 $\mu$m. We present PSF-fitting photometry of archival \textit{Spitzer}/IRAC images for 11 young, low-mass ($M\sim0.044$--0.88 $M_{\odot}$; M7.5--K3.5) members of three nearby star-forming regions (Chameleon, Taurus, and Upper Scorpius; $d\sim$150 pc; $\tau\sim$1--10 Myr) that host confirmed or candidate faint companions at $\rho = 1\farcs68-7\farcs31$. We recover all system primaries, six confirmed, and two candidate low-mass companions in our sample. We also measure non-photospheric $[3.6]-[8.0]$ colors for three of the system primaries, four of the confirmed companions, and one candidate companion, signifying the presence of circumstellar or circum(sub)stellar disks. We furthermore report the confirmation of a $\rho=4\farcs66$ (540 au) companion to [SCH06] J0359+2009 which was previously identified as a candidate via imaging over five years ago, but was not studied further. Based on its brightness ($M_{[3.6]}=8.53$ mag), we infer the companion mass to be $M=20\pm5$ $M_\mathrm{Jup}$ given the primary's model-derived age of 10 Myr. Our framework is sensitive to companions with masses less than 10 $M_\mathrm{Jup}$ at separations of $\rho = 300$ AU in nearby star-forming regions, opening up a new regime of parameter space that has yet to be studied in detail, discovering planetary-mass companions in their birth environments and revealing their circum(sub)stellar disks.
\end{abstract}
\keywords{binaries:general -- methods:observational -- stars:low-mass -- techniques:photometric}
\maketitle

\section{Introduction}
The vast majority of exoplanets have been discovered via the radial velocity \citep{wright11} or transit \citep{thompson18} planet-search methods but the detailed study of their atmospheres and assembly is hindered by their close proximity to bright stellar hosts. Advancing techniques in high-contrast imaging are enabling the detailed study of gas giant planets on wider orbits, providing insight into their formation conditions, atmospheric composition, and circumplanetary environments. Direct imaging surveys typically target nearby young stellar moving groups and associations because contrast ratios between companion and host star is lowest at early ages. These searches have uncovered an interesting population of planetary-mass companions ($\lesssim$20 $M_{\mathrm{Jup}}$; hereafter PMCs) located at wide separations ($>$100 au) from their host stars, such as 1RXSJ1609B (8 $M_{\mathrm{Jup}}$, 330 au; \citealp{lafreniere08}), GSC 06214--00210B (14 $M_{Jup}$, 330 au; \citealp{ireland11}), and HD 106906 b (11 $M_{\mathrm{Jup}}$; 650 au; \citealp{bailey14}). Yet only eight confirmed planets and 15 candidate planets or PMCs on orbits with semi-major axes greater than 100 au have been directly-imaged \citep{bowler16} indicating PMC demographics and occurrence rates suffer from inadequate statistics.

While the deuterium-burning limit at $\approx$13 $M_{\mathrm{Jup}}$ is commonly used as the boundary between what is a giant planet or brown dwarf, that mass definition is still a matter of debate (e.g., \citealp{schneider11,hatzes15,bowler16}). In this work we adopt a definition of ``planetary-mass" $\le$ 20 $M_{\mathrm{Jup}}$ recognizing that the giant planet and brown dwarf companion mass functions overlap between 5--30 $M_{\mathrm{Jup}}$ \citep{wagner19}.

The mere existence of PMCs suggests that such objects are a viable, albeit rare, outcome of star and planet formation. A recent meta-analysis of imaging surveys by \citet{bowler16} found the occurrence rate of planets with masses between 5 and 13 $M_\mathrm{Jup}$ to be $<$2.1\%. \citet{ireland11} found an occurrence rate of planets between 6 and 20 $M_\mathrm{Jup}$ to be 4\% around solar-type stars, suggesting a slightly more optimistic frequency. However, star and planet formation models are inadequately predicting the occurrence rate of wide-orbit PMC systems, and it is still unclear whether they represent the low-mass extreme of the stellar binary formation process, or instead are the end result of high-mass and wide-orbit planet formation process. Opacity-limited fragmentation during the collapse of a molecular cloud can form bodies of the characteristic PMC mass (e.g., \citealp{low76,silk77,boss88,boyd05}) but if the fragment forms before it is isolated from the exhaustion of the circumstellar envelope ($<$0.1 Myr; \citealp{bodenheimer01}), it will accrete enough material to become an object of brown dwarf mass or higher (e.g., \citealp{bate05,tomida13}). Gravitational instability (GI) models might be a plausible alternative formation channel for PMCs should the circumstellar disks they form within be atypically massive and cold enough (e.g., \citealp{kratter10,boss11,vorobyov13}) to undergo fragmentation, though the epoch of \textit{in situ} formation would also need to occur during the Class 0/I stage (0.1 to 0.5 Myr) just as the envelope is exhausted to prevent any further accretion (e.g., \citealp{kratter10,dodson09}). Disk fragmentation could occur later during the Class II stage, but those disks would also need to be unusually massive (e.g., \citealp{andrews13,vorobyov13}).

From the planet formation perspective, the large separations between PMCs and their host stars disfavor core accretion models since the $>$100 Myr timescale required to assemble a core is much longer than the $\sim$3--5 Myr disk lifetime \citep{pollack96}. Dynamical processes like ejection could explain the origins of the PMC population. However, scatterers have not been found \citep{bryan16,pearce19} and measurements of PMC orbital arcs suggest they have low eccentricities \citep{ginski14,schwarz16}, precluding ejection.

\begin{deluxetable*}{lllrrrc}[t!]
\tablecaption{Primary Properties for Test Case Systems}
\tablehead{\colhead{2MASS} & \colhead{Other Name} & \colhead{SpT} & \colhead{$K_s$} & \colhead{$W1$} & \colhead{$W2$} & \colhead{Ref.} \\
& & & \colhead{(mag)} & }
\startdata
J03590986+2009361	& [SCH06] J0359099+2009362		& M4.75	& 12.53	& 12.22	& 11.96	& 1\\
J04233539+2503026	& FU Tau A					& M7.25	& 9.32	& 8.60	& 7.82	& 2\\
J04292971+2616532	& FW Tau AB					& M4		& 9.39	& 9.20	& 8.93	& 3\\
J04554970+3019400	&							& M6		& 11.86	& 11.50	& 11.17	& 1\\
J05373850+2428517	& [SCH06] J0537385+2428518		& M5.25	& 10.78	& 10.60	& 10.34	& 1\\
J11075588--7727257	& CHXR 28 Aa,Ab				& K6		& 7.69	& 6.99	& 7.50	& 4\\
J11122441--7637064	& Sz 41 A						& K3.5	& 8.00	& 6.61	& 6.20	& 4\\
J16101918--2502301	& USco 1610--2502 A			& M1		& 8.36	& 8.24	& 8.20	& 5,6\\
J16103196--1913062	& USco 1610--1913 A			& K7		& 8.99	& 8.74	& 8.67	& 5,6\\
J16111711--2217173		& [SCH06] J16111711--22171749	& M7.5	& 13.25	& 13.07	& 12.74	& 1\\
J16151116--2420153	& [SCH06] J16151115--24201556	& M6		& 13.17	& 12.98	& 12.76	& 1
\enddata
\tablecomments{The primaries for FW Tau and CHXR 28 are actually close binaries but treated as single stars because they are spatially unresolved at the angular scale of \textit{Spitzer}/IRAC observations. Primary properties obtained from the following references: (1) \citet{kraus12}; (2) \citet{luhman09}; (3) \citet{kraus14}; (4) \citet{lafreniere08}; (5) \citet{kraus09b}; (6) \citet{aller13}.}
\label{prim_tab}
\end{deluxetable*}

Thermal disk emission and accretion signatures have been detected from wide-orbit PMCs and substellar companions at a variety of wavelengths. Early searches for ultra-low mass companions identified objects potentially harboring disks through their near-infrared colors. \citet{ireland11} speculated that GSC 06214--00219 b could have a disk based on its red $K^{\prime}-L^{\prime}$ color. Spectroscopic follow-up of the system by \citet{bowler11} confirmed the presence of a disk by observing strong Pa$\beta$ emission. \citet{kraus14} found the majority of their PMC sample to have redder $K^{\prime}-L^{\prime}$ than free-floating young objects which could indicate the ubiquity of disks around wide low-mass companions. Additional signs of outflows and accretion signaling disks surrounding wide-orbit PMCs have been observed with H$\alpha$ line emission and continuum excess in the optical (e.g., \citealp{bowler14,zhou14}). Determining the disk properties of wide-orbit PMCs will help provide additional constraints on their formation pathway. The presence of substellar disks suggest ejection is not a plausible formation mechanism for PMCs because the the scattering event would likely disrupt the disk \citep{bowler11}.

The \textit{Spitzer} mission has obtained a wealth of wide and deep imaging of nearby molecular clouds and cores, including complete Infrared Array Camera (IRAC; \citealt{fazio04}) maps of every major star-forming region within 300 pc (e.g., \citealt{evans09}) across its four channels (3.6, 4.5, 5.8, and 8 $\mu m$; Ch 1,..., Ch 4). The extensive \textit{Spitzer}/IRAC data set of nearby star-forming regions and associations has great potential to be mined for undiscovered wide companions to stars. \textit{Spitzer}/IRAC is capable of resolving companion projected separations of $1\farcs7$ to $2\farcs0$, corresponding to 240 to 290 au at the distances of Taurus or Upper Scorpius ($\sim$145 pc; \citealt{torres09}, \citealt{deZeeuw99}) and sensitive to the photospheres of proto-brown dwarfs and protoplanets ($M_{lim}$ = 1 $M_{\mathrm{Jup}}$ at 1 Myr and 2 $M_{\mathrm{Jup}}$ at 5 Myr; \citealt{chabrier00b}). \textit{Spitzer}'s limits are even deeper for hosts of circumplanetary disks, which add substantial infrared excesses.

In this paper, we report the results of a pilot study to recover and to measure the mid-infrared photometry of confirmed and candidate low-mass companions of young stars in \textit{Spitzer}/IRAC imaging and to determine whether they host circum(sub)stellar disks. In Sections 2 and 3, we describe our target sample and the archival \textit{Spitzer}/IRAC observations used. We describe our MCMC based PSF-fitting routine in Section 4. In Section 5, we present the results of our analysis of the IRAC images and highlight interesting individual systems. Finally in Section 6, we discuss the performance of our PSF-fitting routine.

\section{Target Sample}
Our target sample is built from systems with confirmed or candidate low-mass companions previously discovered in the star-forming regions of Chameleon I (179 pc, 2--3 Myr; \citealt{voirin18}, \citealt{luhman04}), Taurus-Auriga ($145\pm15$ pc, 1--2 Myr; \citealt{torres09}, \citealt{kraus09a}), and Upper Scorpius (145 pc, 5--10 Myr; \citealt{deZeeuw99}, \citealt{preibisch02}, \citealt{pecaut12}) from the multiplicity surveys of \citet{lafreniere08} and \citet{kraus12}. We also include in our sample a small number of systems observed by \citet{kraus12} that might be a part of an older distributed population of Taurus (5--10 Myr; \citealt{wichmann96}, \citealt{slesnick06}, \citealt{kraus17}). These young regions are compelling targets because they offer increased sensitivity to PMCs retaining residual heat from formation. There is also a high likelihood of finding companions harboring disks, which would add substantial infrared excess.

From this larger sample, we then identify systems that also have archival \textit{Spitzer}/Infrared Array Camera (IRAC; \citealt{fazio04}) observations available. In-flight full width at half-maximums (FWHMs) for the IRAC PSF are $1\farcs66$, $1\farcs72$, $1\farcs88$, and $1\farcs98$ in each respective channel, corresponding to resolvable companion separations above 240 au at the distances of Taurus or Upper Scorpius. IRAC is also sensitive enough to detect photospheres of proto-brown dwarfs and protoplanets ($M_{\mathrm{lim}}$ = 1 $M_{\mathrm{Jup}}$ at 1 Myr and 2 $M_{\mathrm{Jup}}$ at 5 Myr; \citealt{chabrier00b}) at wide separations in the background-limited regime. Our sample was intentionally constructed to span a range of primary brightness, contrast, and projected separation to test our ability to recover astrophysical sources and sufficiently probe those axes of parameter space. Most of the systems in the target sample have high-precision astrometry gathered from previous adaptive optics (AO) imaging. We used these measurements to experiment with the effects of using informative priors.

\begin{deluxetable*}{llcccr}
\tablecaption{Properties of Test Case Confirmed and Candidate Companions}
\tablehead{\colhead{2MASS} & \colhead{Other Name} & \colhead{Separation} & \colhead{Position Angle} & \colhead{$\Delta K_s$} & \colhead{Ref.} \\
& & \colhead{(arcsec)} & \colhead{(deg)} & \colhead{(mag)} & }
\startdata
J03590986+2009361 B	& [SCH06] J0359099+2009362	B	& $4.660\pm0.005$	& $264.275\pm0.003$	& $1.965\pm0.005$	& 1 \\
J04233539+2503026 B	& FU Tau B					& $5.72\pm0.10$	& $123.2\pm1.0$		& $4.01\pm0.10$	& 2\\
J04292971+2616532 C	& FW Tau C					& $2.295\pm0.003$	& $295.0\pm0.5$		& $5.93\pm0.04$	& 3\\
J04554970+3019400 c1	&							& $7.313\pm0.007$	& $129.15\pm0.02$		& $1.77\pm0.05$	& 1\\
J05373850+2428517 c1	& [SCH06] J0537385+2428518	c1	& $1.684\pm0.008$	& $152.84\pm0.14$		& $7.27\pm0.13$	& 1\\
J11075588--7727257 B	& CHXR 28 B					& $1.818\pm0.003$	& $115.9\pm0.1$		& $0.32\pm0.04$	& 4\\
J11122441--7637064 B	& Sz 41 B						& $1.977\pm0.001$	& $162.5\pm0.1$		& $2.35\pm0.03$	& 4\\
J16101918--2502301 B	& USco 1610--2502 B			& $4.896\pm0.002$	& $241.24\pm0.02$		& $2.90\pm0.05$	& 5,6\\
J16103196--1913062 B	& USco 1610--1913 B			& $5.820\pm0.009$	& $114.01\pm0.10$		& $3.83\pm0.05$	& 5,6\\
J16111711--2217173	 c1	& [SCH06] J16111711--22171749 c1	& $4.207\pm0.004$	& $344.41\pm0.02$		& $5.66\pm0.05$	& 1\\
J16151116--2420153 c1	& [SCH06] J16151115--24201556 c1	& $5.100\pm0.005$	& $141.03\pm0.01$		& $4.74\pm0.02$	& 1
\enddata
\tablecomments{An object is labeled as ``c\#" to reflect that they are unconfirmed candidate companions. System properties obtained from the following references: (1) \citet{kraus12}; (2) \citet{luhman09}; (3) \citet{kraus14}; (4) \citet{lafreniere08}; (5) \citet{kraus09b}; (6) \citet{aller13}.}
\label{comp_tab}
\end{deluxetable*}

\citet{lafreniere08} examined 126 stars ranging in mass from $\sim$0.1--3$M_\odot$ in Chameleon I. Using the ESO Very Large Telescope AO imaging system, they found 30 binary and six triple systems. We chose CHXR 28 and Sz 41 from that study for our target sample. \citet{kraus12} used Keck laser guide star AO to study 78 stars in Taurus-Auriga and Upper Scorpius, finding 45 candidate companions. We chose five candidate systems from that survey for our sample: 2MASS J03590986+2009361, 2MASS J04554970+3019400, 2MASS J05373850+2428517, 2MASS J16111711--2217173, and 2MASS J16151116--2420153. We also incorporate four more systems into our target sample not studied in the aforementioned surveys: FU Tau AB \citep{luhman09}, FW Tau \citep{kraus14}, 2MASS J16101918--2502301 \citep{aller13}, and 2MASS J16103196--1913062 \citep{aller13}.

The primary spectral types for the 11 systems in our target sample range from K3.5 to M7.5 and the primary $K_s$-band magnitudes range from 7.69 mag to 13.25 mag. The projected separations and $K_s$-band contrasts of the candidate and confirmed companions range from $1\farcs684$ to $7\farcs313$ and $0.32$ to $7.27$ mag, respectively. Four of the confirmed low-mass companions in these systems are known to have disks; FU Tau B, FW Tau C, Sz 41 B, and USco 1610--2502 B. The combined target primary properties are given in Table \ref{prim_tab} and system properties in Table \ref{comp_tab}.

\section{Observations}

All targets were observed by the \textit{Spitzer Space Telescope} \citep{werner04} with the Infrared Array Camera (IRAC; \citealt{fazio04}) during the cryogenic phase of the mission. IRAC operates with four filters in the mid-infrared; 3.6, 4.5, 5.8, and 8.0 $\mu$m. The IRAC detector has 256$\times$256 pixels with a pixel scale of $1\farcs22$.

Observations of the 11 members of our target sample appear in ten different sets of IRAC data, with exposure times of 0.4 s, 1.0 s, 10.4 s, 26.8 s, and 96.8 s. Almost all targets had data taken across the four IRAC channels, but USco 1610--1913 did not have any Channel 1 or Channel 3 observations. Specific details about the \textit{Spitzer}/IRAC programs and data products used are listed in Table \ref{files_tab}.

We work with IRAC's cryogenic-phase corrected basic calibrated data (CBCD) and uncertainty (CBUNC) files. Mosaics were not used due to the complicated behavior of the IRAC PSF (see Data Analysis section). We use images with exposure times of 10.4 s or less to avoid saturation.

All data were reduced with the \textit{Spitzer} Science Center software pipeline version S18.25.0. No further reduction or processing of the images was performed.

\section{Data Analysis}
\begin{figure*}
\centering
\includegraphics[trim={0.4in 0.75in 0.25in 0.75in},width=1.0\textwidth]{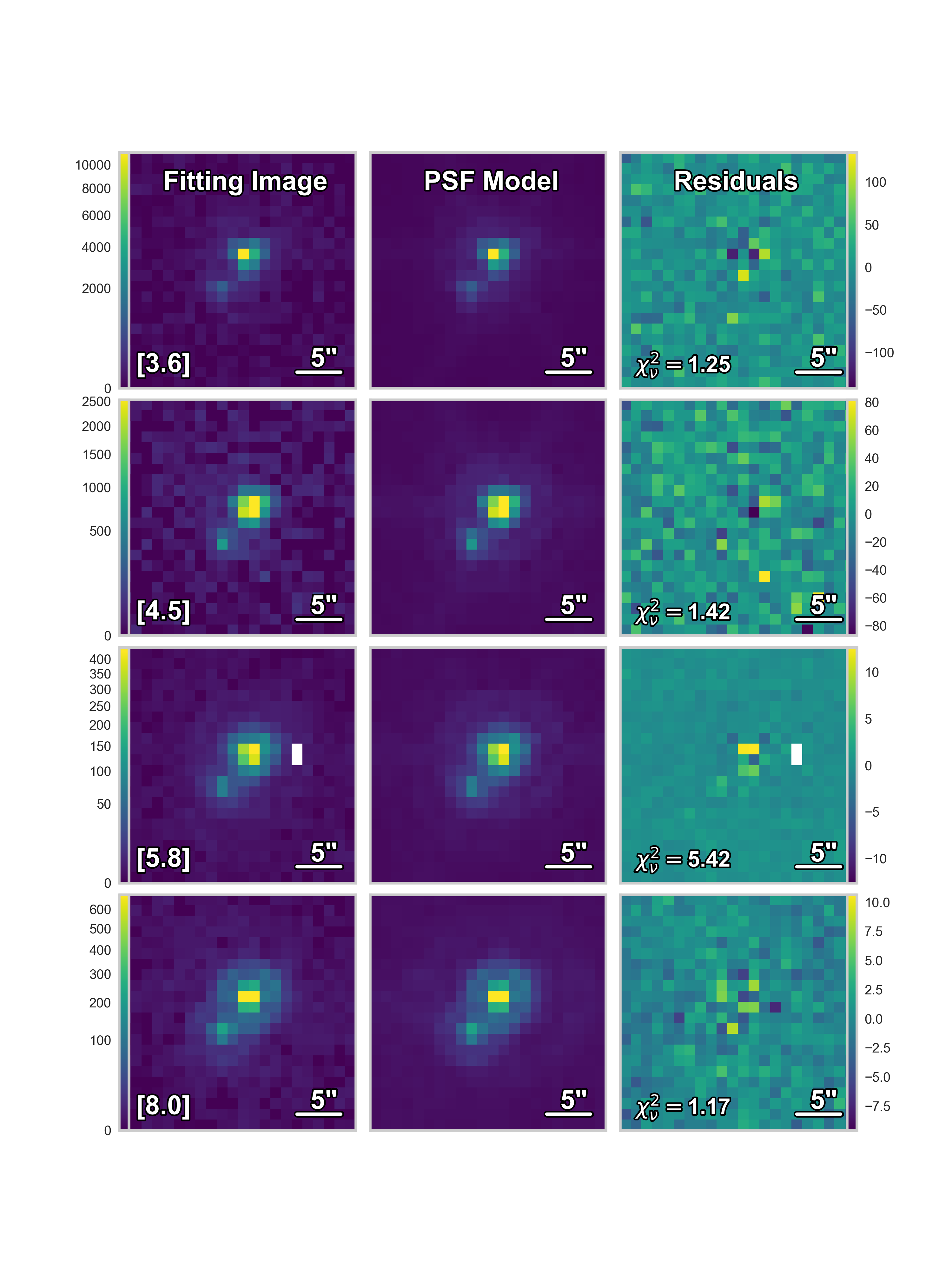}
\caption{Images of USco 1610--2502 at four stages of the PSF-fitting pipeline. The pipeline has been generalized to perform PSF-subtraction across all four channels (rows) should images be available. {\bfseries Column 1:} The background-subtracted fitting image in units of Data Number per second (DN/s) presented with a square root stretch. {\bfseries Column 2:} The best-fit PSF model in units of DN/s and also presented with a square root stretch. {\bfseries Column 3:} The residuals left over after the PSF model is subtracted from the background subtracted fitting image, plotted with a linear color scale. Cosmic rays (for example, white pixels in row 3, column 1) that fall near the system are masked prior to PSF-fitting. Note that the remaining residuals appear similar to the noise near the primary star, indicating a good fit.}
\label{USco1610-2502}
\end{figure*}

\begin{deluxetable*}{lcclccclrl}
\tablecaption{Target Sample \textit{Spitzer}/IRAC Observations}
\tablehead{\colhead{2MASS}	& \multicolumn{4}{c}{No.\ of Frames}		& \colhead{$T_{\mathrm{exp}}$}	& \colhead{AOR}	& \colhead{Date}	& \colhead{PID}	& \colhead{PI}	\\ \cline{2-5} \\ \multicolumn{10}{c}{\vspace{-0.5cm}}\\
				& Ch 1	& Ch 2	& Ch 3	& Ch 4	& \colhead{(s)}					&		& \colhead{(UT)}	&			&				}
\startdata
J03590986+2009361	& 4		& 4		& 4		& 4	& 0.4, 10.4	& 26469888	& 2009 Mar 20	& 50584	& D. Padgett	\\
J04233539+2503026	& 		& 3 		& 		& 3	& 0.4, 10.4	& 3964672	& 2005 Feb 23	& 37		& G. Fazio	\\
					& 2 		& 1 		& 2 		& 1	& 0.4, 10.4	& 19028224	& 2007 Mar 30	& 30816	& D. Padgett	\\
 					& 1 		& 1 		& 1 		& 1	& 0.4, 10.4	& 19028480	&			&		&			\\
J04292971+2616532	& 3 		& 3 		& 3 		& 3	& 0.4, 10.4	& 3963392	& 2004 Mar 07	& 37		& G. Fazio	\\
					& 1 		& 1 		& 2 		& 1	& 0.4, 10.4	& 11232256	& 2005 Feb 23	& 3584	& D. Padgett	\\
					& 1 		& 1 		& 1 		& 1	& 0.4, 10.4	& 11236096	& 2005 Feb 24	&		&			\\
					& 8 		& 8 		& 8 		& 8	& 1.0			& 14609920	& 2006 Mar 24	& 20386	& P. Myers	\\
					& 8 		& 8 		& 8 		& 8	& 1.0			& 14610176	& 2006 Mar 25	&		&			\\
J04554970+3019400	& 2 		& 2 		& 1 		& 2	& 0.4, 10.4	& 3965696	& 2004 Feb 14	& 37		& G. Fazio	\\
					& 1 		& 		& 1 		&	& 0.4			& 12663552	& 2005 Feb 20	&		&			\\
					& 6 		& 3 		& 6 		& 3	& 10.4		&			&			&		&			\\
					& 		& 1 		& 		& 1	& 0.4, 10.4	& 26476544	& 2008 Nov 01 	& 50584	& D. Padgett	\\
J05373850+2428517 	& 4 		& 4 		& 4 		& 4	& 0.4, 10.4	& 26478336	& 2008 Oct 31	& 50584	& D. Padgett	\\
J11075588--7727257 	& 6 		& 3 		& 7 		& 3	& 0.4, 10.4	& 3960320	& 2004 Jun 10	& 37		& G. Fazio	\\			
J11122441--7637064	& 3 		& 6 		& 3 		& 7	& 0.4, 10.4	& 3651328	& 2004 Jul 04	& 6		& G. Fazio	\\				
					& 		& 1 		& 		& 1	& 0.4, 10.4	& 5662976	& 2004 Jul 21	& 173	& N. Evans	\\
J16101918--2502301 	& 2 		& 2 		& 2 		& 2	& 0.4, 10.4	& 5670912	& 2004 Aug 12	& 173	& N. Evans	\\
J16103196--1913062	& 		& 9 		& 		& 9	& 1.2			& 13868288	& 2005 Sep 15	& 20069	& J. Carpenter	\\
					& 		& 8 		& 		& 8	& 10.4		& 13874944	& 2005 Aug 23	&		&			\\
J16111711--2217173		& 		& 1 		& 		& 1	& 0.4			& 15843072	& 2005 Aug 24	& 20103	& L. Hillenbrand\\
					& 5 		& 5 		& 5 		& 5	& 10.4		&			&			&		&			\\
J16151116--2420153	& 		& 1 		& 		& 1	& 0.4 		& 15837440	& 2005 Aug 24	& 20103	& L. Hillenbrand\\
					& 5 		& 5 		& 5 		& 5	& 10.4		&			&			&		&									
\enddata
\label{files_tab}
\end{deluxetable*}

\begin{figure*}
\centering
\begin{minipage}{0.45\textwidth}
\includegraphics[width=1.0\textwidth]{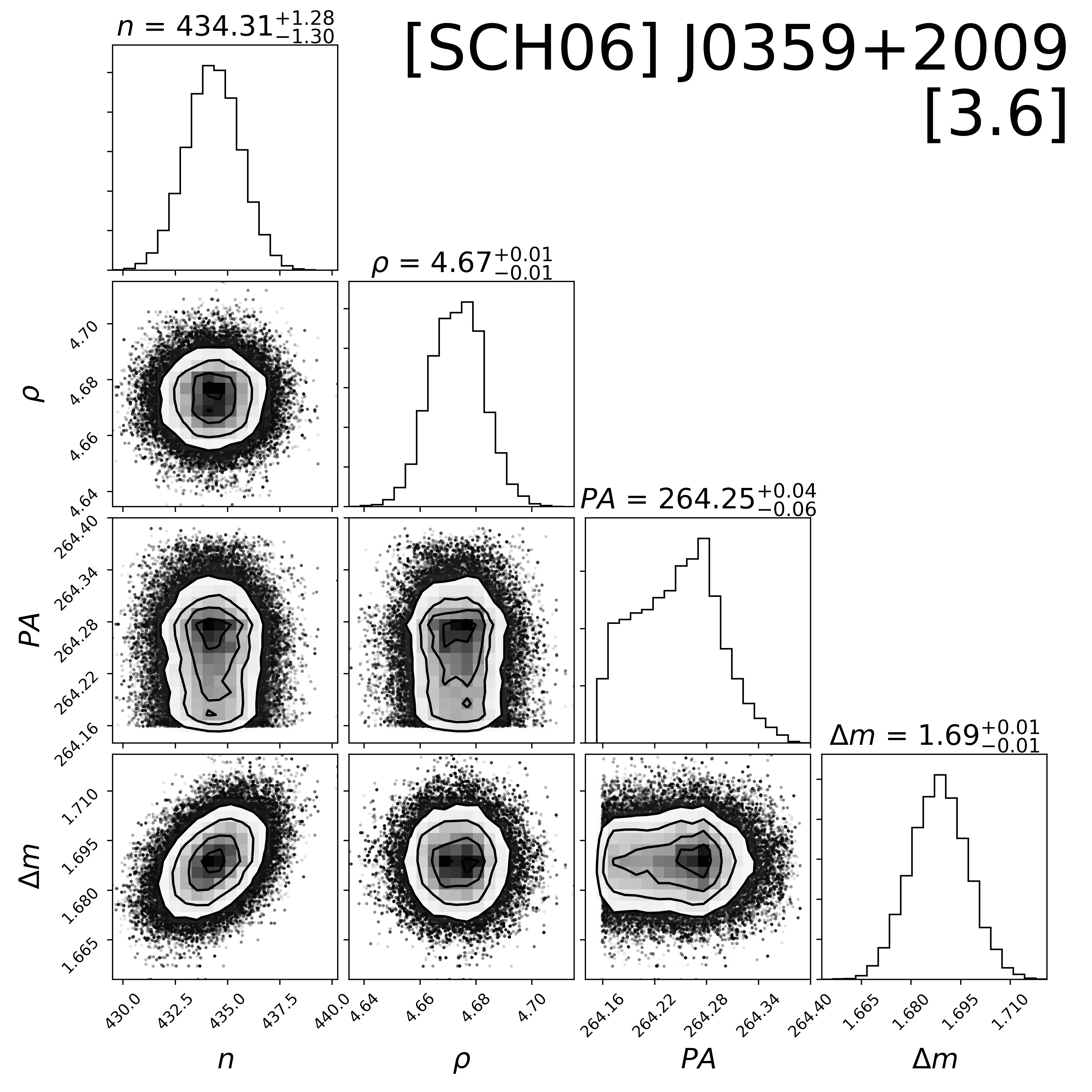}
\end{minipage}
\begin{minipage}{0.45\textwidth}
\includegraphics[width=1.0\textwidth]{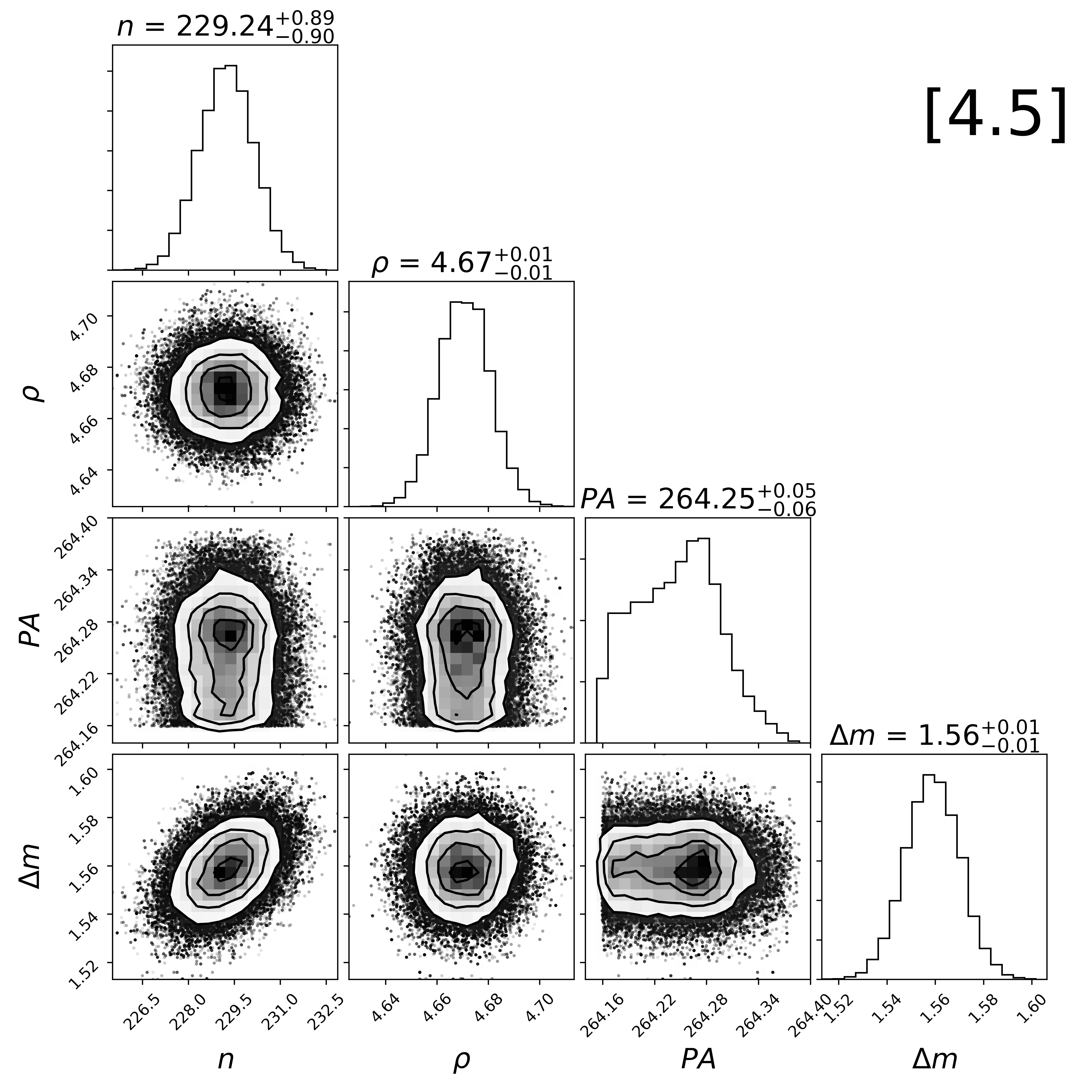}
\end{minipage}
\begin{minipage}{0.45\textwidth}
\includegraphics[width=1.0\textwidth]{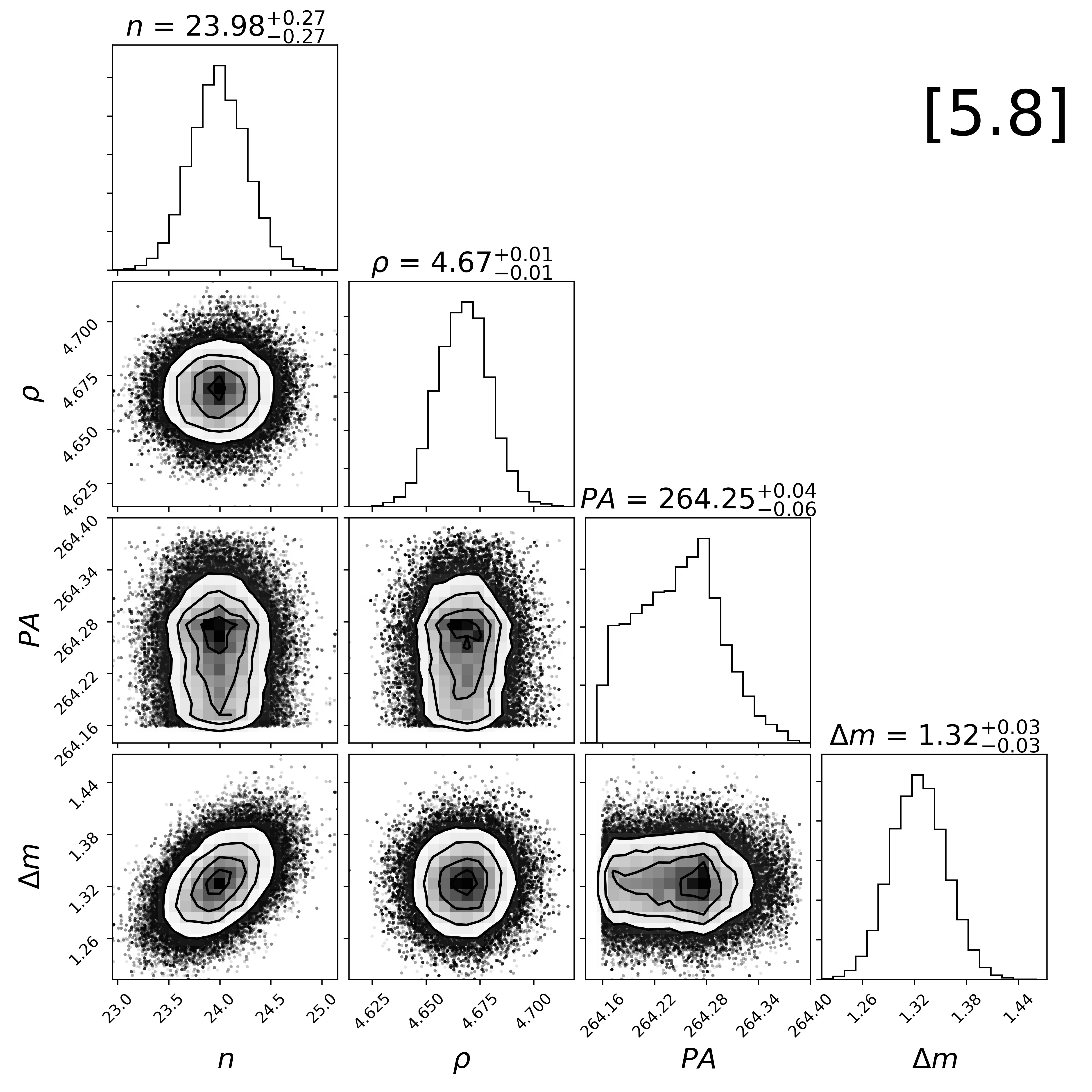}
\end{minipage}
\begin{minipage}{0.45\textwidth}
\includegraphics[width=1.0\textwidth]{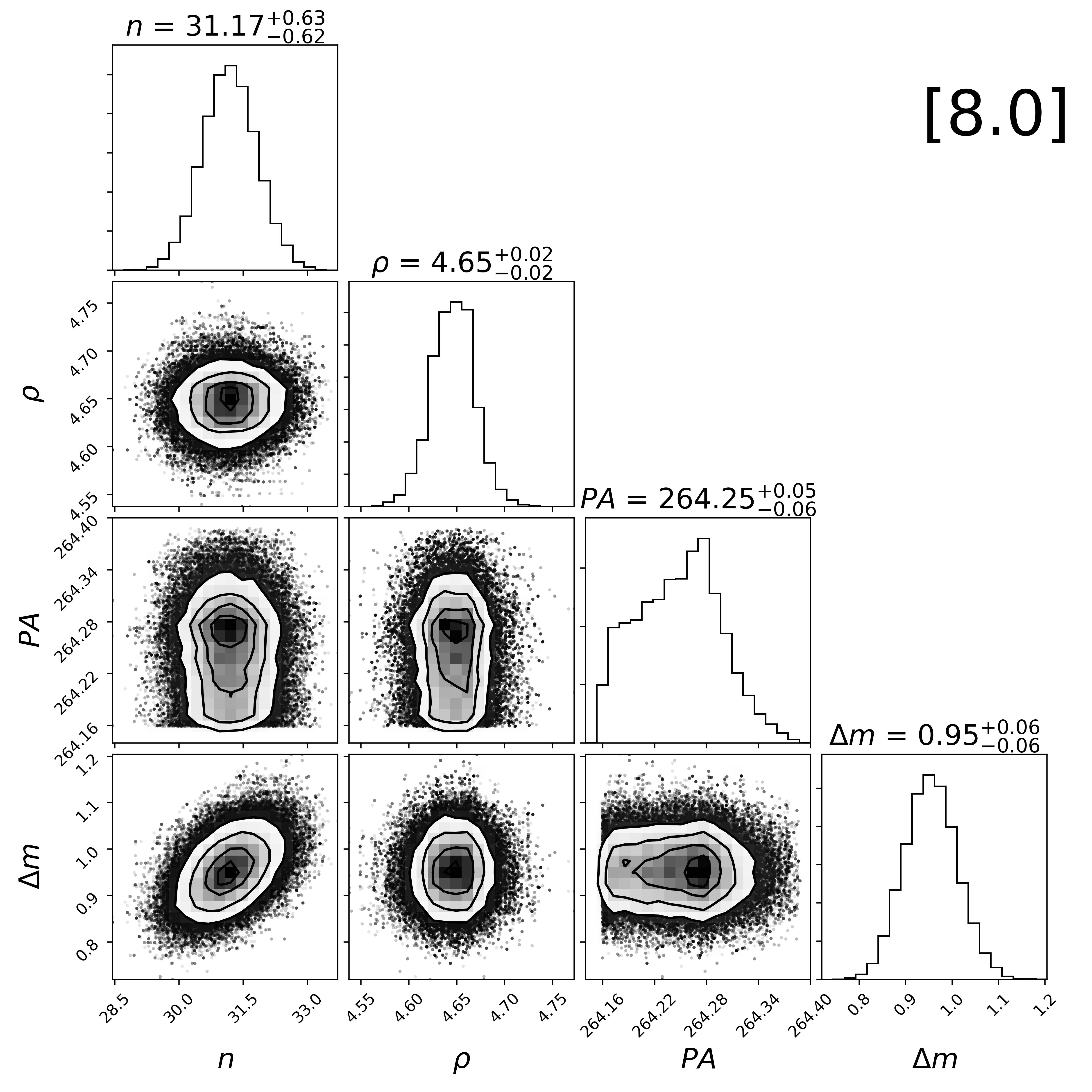}
\end{minipage}
\caption{Posterior probability distributions of the four system-specific parameters fit to the images of [SCH06] J0359+2009, an example system that had its candidate companion detected by our pipeline. No strong covariances are present due to the strong prior on the astrometry from previous adaptive optics imaging. The asymmetric appearance of the PA marginalized posterior probability density results from the pixelation of the IRAC PSF models.}
\label{pdfs_schj0359}
\end{figure*}

\begin{figure*}
\centering
\begin{minipage}{0.45\textwidth}
\includegraphics[width=1.0\textwidth]{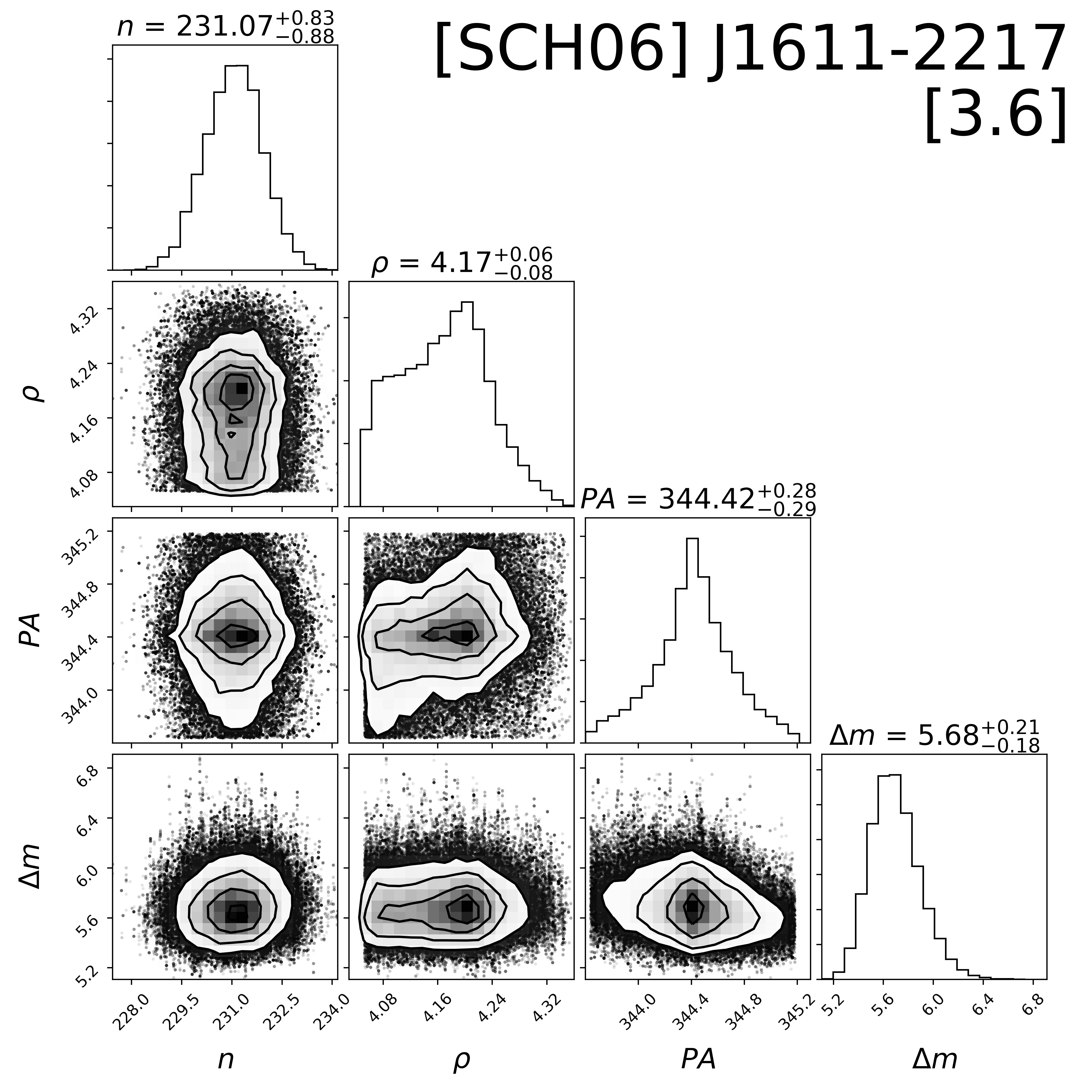}
\end{minipage}
\begin{minipage}{0.45\textwidth}
\includegraphics[width=1.0\textwidth]{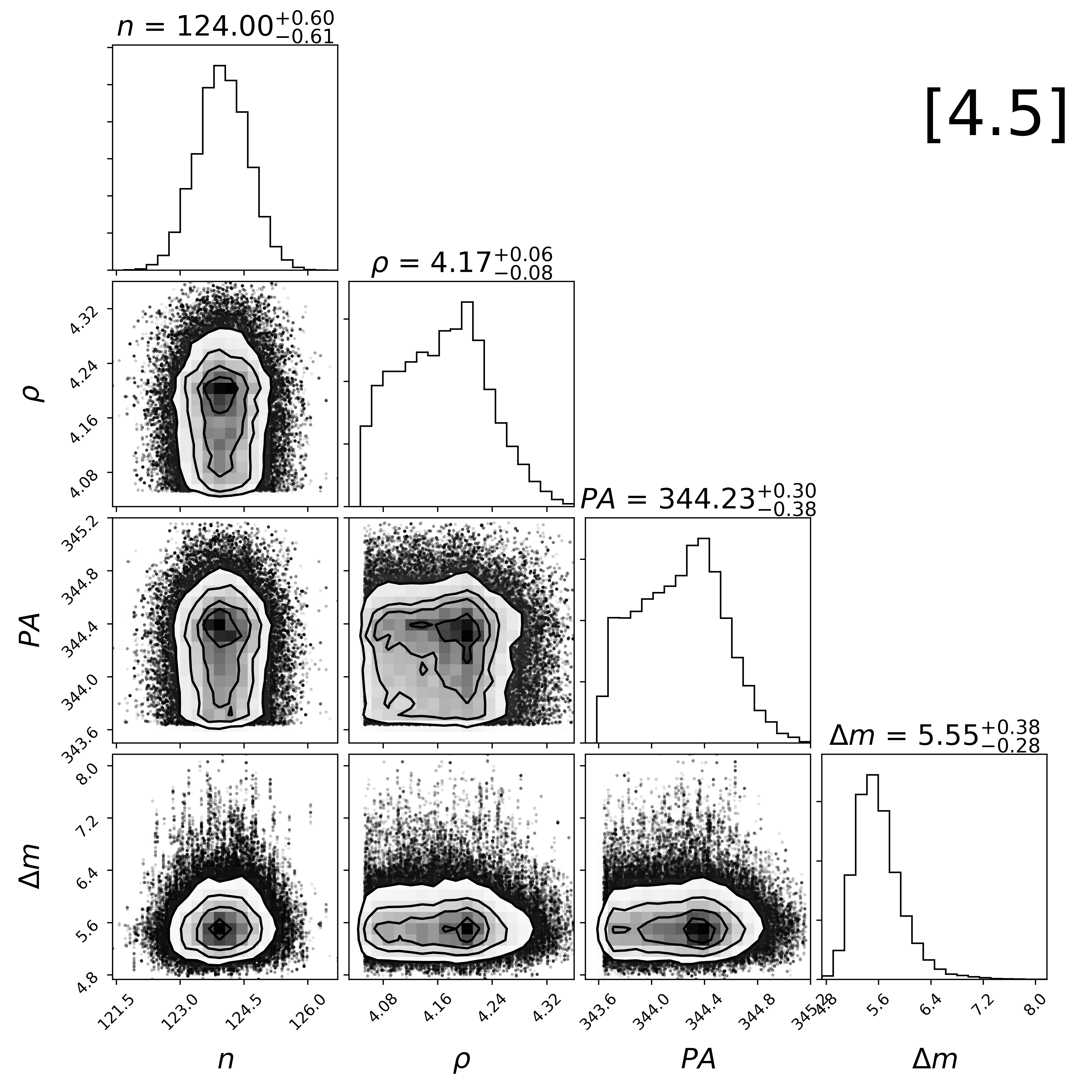}
\end{minipage}
\begin{minipage}{0.45\textwidth}
\includegraphics[width=1.0\textwidth]{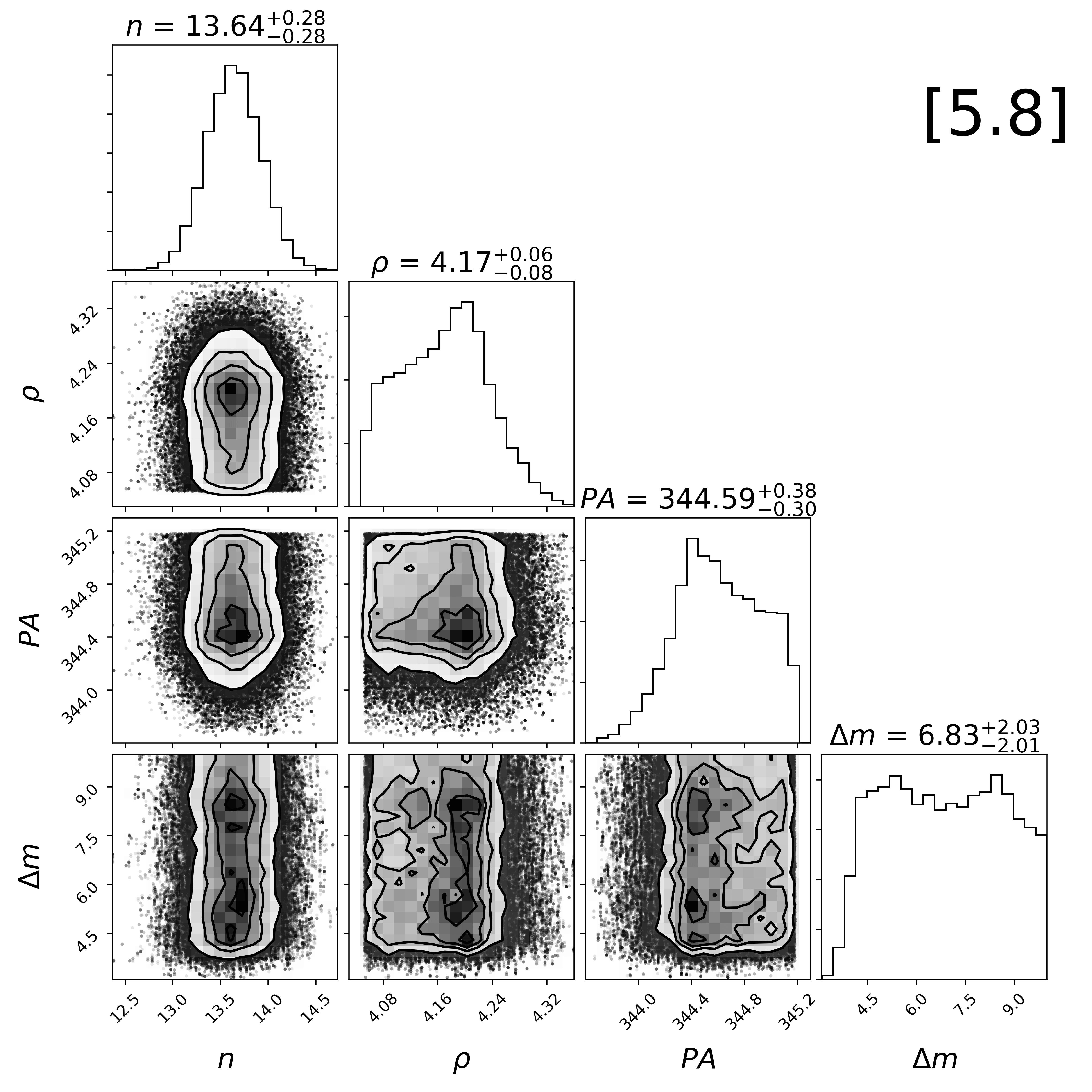}
\end{minipage}
\begin{minipage}{0.45\textwidth}
\includegraphics[width=1.0\textwidth]{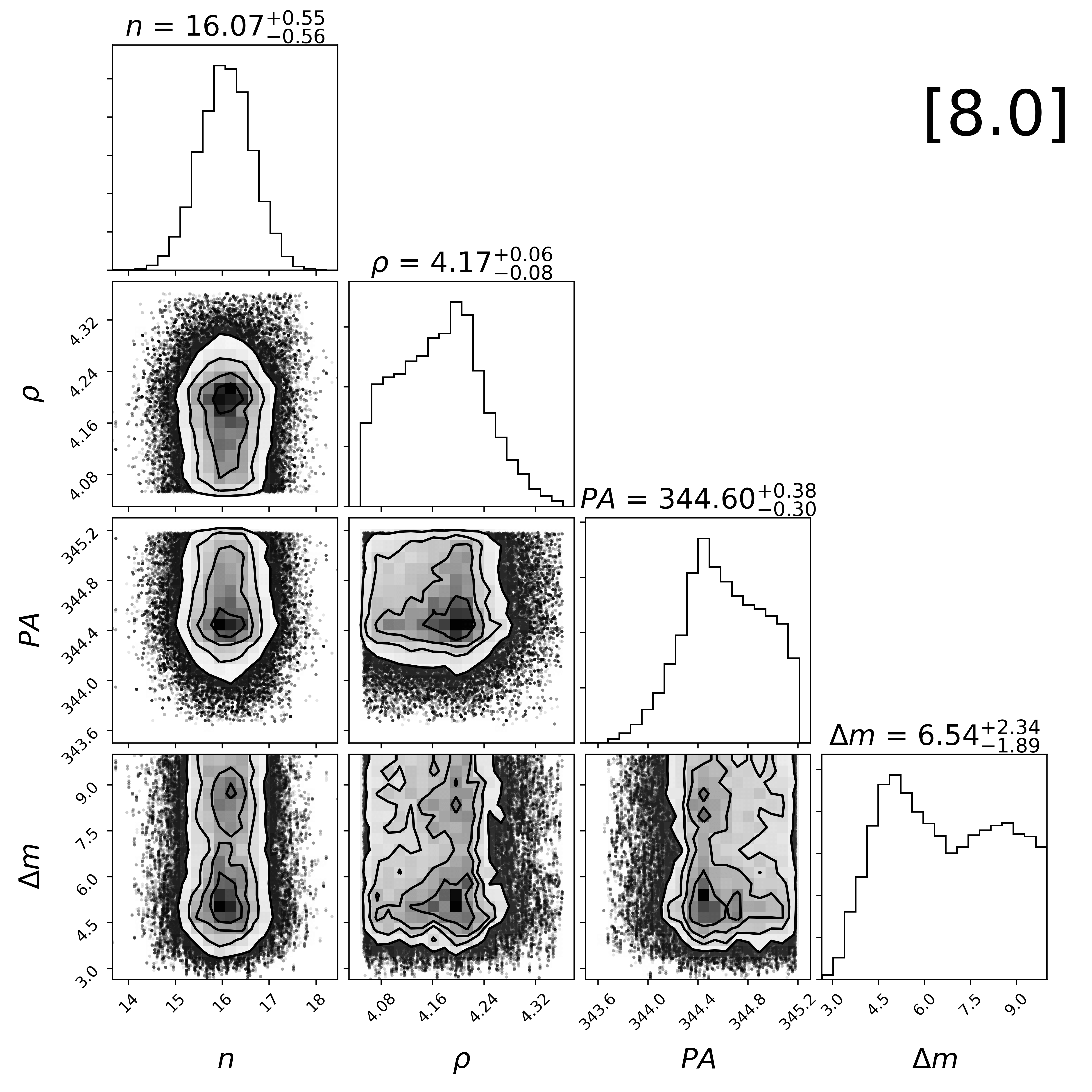}
\end{minipage}
\caption{Posterior probability distributions of the four system-specific parameters fit to the images of [SCH06] J1611-2217, an example system where the candidate companion was not detected by our pipeline. The projected separation and PA marginalized posterior probability densities are still well-constrained because of the strong prior on the astrometry from previous adaptive optics imaging. The marginalized posterior probability densities for $\Delta m$ in Channels 1 and 2 appear constrained but inspection of stacked residuals images reveal no companion detected above our 3-$\sigma$ detection limit (see Section \ref{dls}). The marginalized posterior probability densities for $\Delta m$ in Channels 3 and 4 are unconstrained.}
\label{pdfs_schj1611}
\end{figure*}

\subsection{The IRAC PSF}
Previous analyses of \textit{Spitzer}/IRAC images have searched for wide-orbit PMC systems, taking advantage of IRAC's well-behaved PSF wings at $>>\lambda/D$. \citet{marengo06} first established that \textit{Spitzer}/IRAC has the ability to detect PMCs in the background-limited regime in their search for companions orbiting $\epsilon$ Eridani. Similar studies of Vega, Fomalhaut, and $\epsilon$ Eridani IRAC images by \citet{janson15} utilized more sophisticated post-processing techniques (i.e., locally-optimized combination of images, principal component analysis, empirical stellar templates) to demonstrate that \textit{Spitzer} was sensitive to PMCs that lie closer to their host stars. \citet{durkan16} reanalyze archival \textit{Spitzer}/IRAC direct imaging surveys of nearby stars to constrain the frequency of giant planets orbiting out to 1000 au. Most recently, \citet{baron18} reported first results from the Wide-orbit Exoplanet search with InfraRed Direct Imaging (WEIRD) survey constraining the occurrence of Jupiter-like companions on orbits between 1000 and 5000 au. These studies focused their investigations on regions closer ($d<100$ pc) and older ($\tau>10$ Myr) than the regions we analyze in this work ($d>100$ pc; $\tau<10$ Myr).

Our framework is probing the IRAC PSF at 1--3 $\lambda/D$, where companion identification is difficult due to it being undersampled at the native $1\farcs22$ pixel scale. To overcome this obstacle when measuring photometry on IRAC images via point source fitting, a PRF (or ``effective PSF") was developed by the \textit{Spitzer} Science team that combined information regarding the IRAC PSF, detector sampling, and intrapixel sensitivity variations. Twenty-five total PRFs are provided that represent 25 different locations ($5\times5$ grid) on the $256\times256$ pixel IRAC detector. Each PRF for a given detector location is $5\times$ oversampled, thus to create the proper PSF for any given position on the detector, one must interpolate the original $5\times5$ grid of PRFs to the centroid position, shift the PRF to the appropriate ``pixel phase," then sample the individual PRF at regular intervals corresponding to single CCD pixel increments.

We use a modified version of {\tt IRACSIM} \footnote{http://dx.doi.org/10.5281/zenodo.46270} \citep{ingalls16}, an Interactive Data Language (IDL) package built to model the pointing, imaging, and Fowler sampling behavior of \textit{Spitzer}/IRAC Channels 1 and 2 in the post-cryogenic mission. \citet{ingalls16} rescale the cryogenic PRFs to accommodate the change in intra-pixel sensitivity during \textit{Spitzer}'s warm mission, using these modified PRFs to define the reference frame of a point source to be modeled in native pixel units. We incorporate these imaging modules into our pipeline by adapting them to use the original core PRFs from the cold mission instead, and adding the ability to generate PSFs for IRAC channels 3 and 4.

\subsection{MCMC PSF Subtraction}
Due to the complicated behavior of the IRAC PSF, we adopt an Markov Chain Monte Carlo (MCMC) formalism to fully explore the resulting posterior probability density function for the system parameters. The PSF model is described by seven parameters, three of which are ``image-specific" (and hence have independent values for each image) and four of which are ``system-specific" (and hence are shared between all images of a target). The three image-specific parameters are x-pixel coordinate of the primary centroid ($x$), y-pixel coordinate of the primary centroid ($y$), and image background ($b$), while the four system-specific parameters are primary peak pixel value ($n$), separation ($\rho$), position angle (PA), and contrast ($\Delta m$). We use the $\chi^2$ goodness-of-fit between the image and PSF model as our likelihood function. The priors associated with each parameter are presented in Table \ref{mcmc_pars}. We use uniform priors for $x$, $y$, $n$, and $\Delta m$. We constrain $x$ and $y$ to be within four pixels of the initial primary centroid estimate. We also constrain $\Delta m$ to be between $-2$ and $10$ mag, allowing for the possibility of a companion being brighter in the IRAC channels than the primary.

The automated PSF-fitting pipeline performs an MCMC analysis by exploring the posterior PDF using the standard Metropolis-Hastings algorithm with Gibbs sampling. We use four walkers (chains) of 56,000 total jumps each, but break this chain up into three separate MCMC fits. The pipeline first runs a 140,000 jump chain fitting all seven parameters for all of the IRAC images available at a given exposure time of a system, then iterates twice between a 80,000 jump chain image-specific parameter fit and a 60,000 jump chain system-specific parameter fit. The first 10 percent of the chains are removed as burn-in. Images were not analyzed if the initial estimate for the target system primary centroid was within 12 pixels of the IRAC CCD edge. In Figure \ref{USco1610-2502} we present example images as a target system is processed through our pipeline. We show example posterior probability distributions for two target systems in Figures \ref{pdfs_schj0359} and \ref{pdfs_schj1611}.

\begin{deluxetable}{cclc}
\tablecaption{MCMC Fit Model Parameters}
\tablehead{Parameter & Symbol & Prior & Constraints}
\startdata
\multicolumn{4}{c}{Image-Specific Fit}\\
\hline \\
\multicolumn{4}{c}{\vspace{-0.5cm}}\\
$x$-centroid & $x$ & uniform & $[x_{0}-4,x_{0}+4]$ \\
$y$-centroid & $y$ & uniform & $[y_{0}-4,y_{0}+4]$ \\
Background & $b$ & normal\tablenotemark{a} \\
\hline \\
\multicolumn{4}{c}{\vspace{-0.5cm}}\\
\multicolumn{4}{c}{System-Specific Fit}\\
\hline \\
\multicolumn{4}{c}{\vspace{-0.5cm}}\\
Peak Pixel Flux & $n$ & uniform & None\\
Projected Separation & $\rho$ & normal\tablenotemark{b} & ... \\
Position Angle & P.A. & normal\tablenotemark{b} & ... \\
Contrast & $\Delta m$ & uniform & $[-2,10]$
\enddata
\tablenotetext{a}{Normal prior on the background was based on the pixel value distribution within a 30-pixel radius of the primary.}
\tablenotetext{b}{Normal prior on these parameters were based on prior adaptive optics imaging results presented Table \ref{comp_tab}.}
\label{mcmc_pars}
\end{deluxetable}

\begin{figure*}
\centering
\includegraphics[trim={0.90in 0.75in 0.75in 0.75in},width=1.0\textwidth]{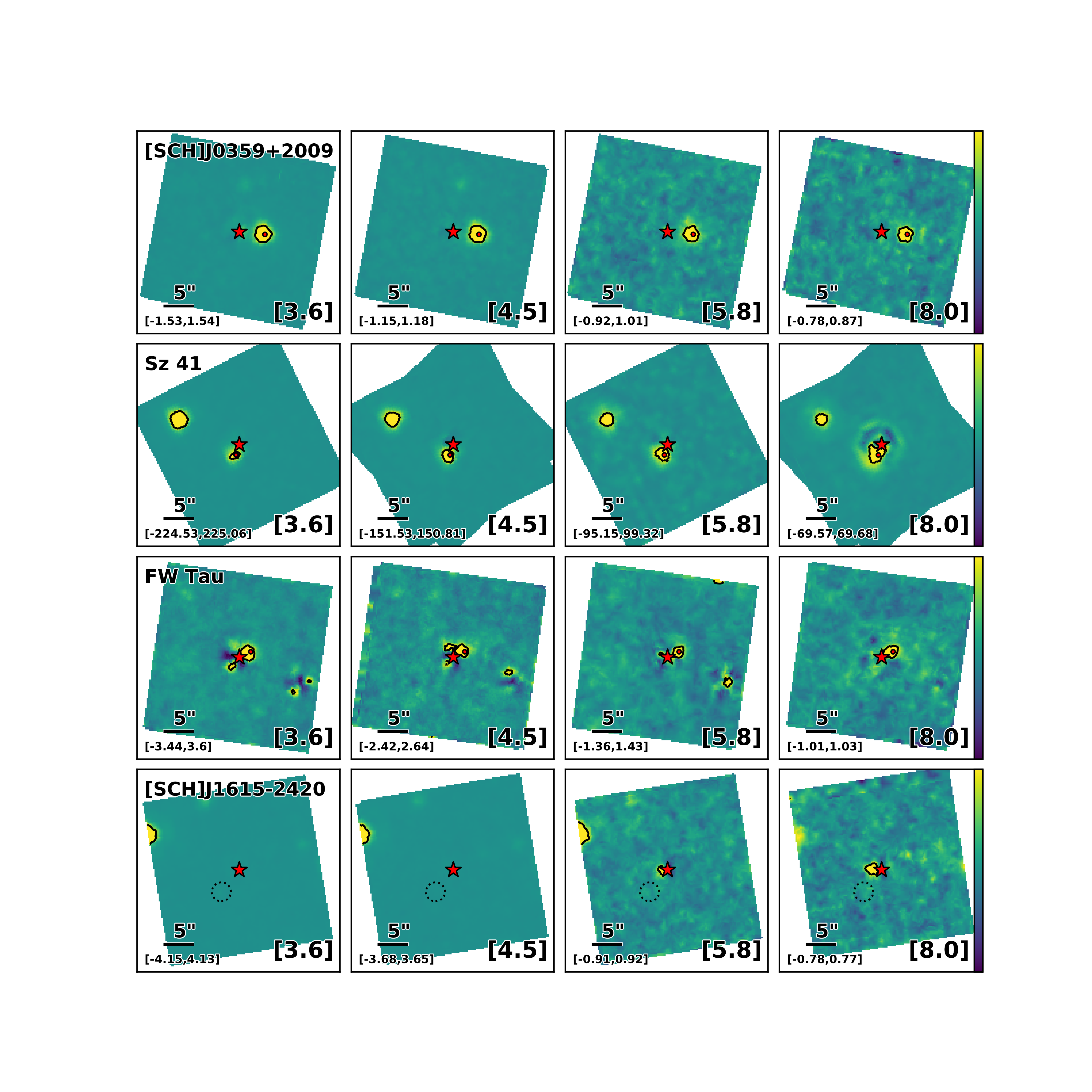}
\caption{Stacked residuals images for [SCH06] J0359+2009, Sz 41, FW Tau, and [SCH06]J1615--2420 (rows) across all four IRAC channels (columns). Images were generated by combining individual residual images after the primary PSF had been subtracted, placing each on a final grid with a pixel scale 5 times smaller than the original IRAC pixel scale of $1\farcs22$, shifting to a common origin, and rotating so that north is up and east is left. The stacked residuals images are displayed with a linear color scale and 5-$\sigma$ contours overlaid. The minimum and maximum pixel value of the color bar are given in the bottom lefthand corner in units of DN/s. A red star denotes the position of the primary while a red or dashed circle denotes the expected position of the companion. {\bfseries First row:} [SCH06] J0359+2009 is an example of a straightforward companion detection. See Section \ref{schj0359} for more details on this system. {\bfseries Second row:} Sz 41 is the brightest member of our sample ($M_{[3.6]}$=0.89) and appears saturated in Channel 4 as evidenced by the ring-like structure in the residuals surrounding the primary location. We are still able to measure photometry for the system that agrees with previous measurements. The other bright object to the upper left of the system is a known background giant. {\bfseries Third row:} Residuals images for FW Tau C, which we are able to resolve across all four IRAC channels for the first time. See Sections \ref{fwtau1} and \ref{fwtau2} for more details on this system. {\bfseries Fourth row:} the candidate companion of [SCH06] J1615--2420 was not detected by our pipeline. A dotted circle is placed at its expected location. The bright object to the upper left is another Upper Sco member.}
\label{stacked_residuals}
\end{figure*}

\section{Results}
By developing the framework to accurately model and subtract off the flux of bright primary stars, we now can take advantage of \textit{Spitzer}'s extraordinary sensitivity to study wide low-mass companions near the diffraction limit. Previous analyses of archival \textit{Spitzer}/IRAC images have searched for wide companions in young moving groups that are closer ($<100$ pc) and older ($>10$ Myr) than the regions represented in our target sample (e.g., \citealp{durkan16,baron18}). Although they were able to take advantage of the well-behaved IRAC PSF wings at $>> \lambda /D$ when searching for wide companions, they do so when the contrast between primary star and companion is more severe. Moreover, some wide PMCs within the more distant star-forming regions harbor disks, so some of our targets should have detectable excesses in the mid-infrared, further improving detectability and offering an opportunity to investigate disk evolution and dispersal in low-mass companions to stars. Our MCMC-based PSF fitter is finally opening up a regime of parameter space that has yet to be studied in detail and will reveal low-mass companions, whether they have disks, and the properties of those disks. 

\subsection{Detections}
Our reprocessing of the IRAC images yielded detection of all 11 system primaries, six confirmed, and two candidate low-mass companions. The targets were processed independently by channel, with a simultaneous fit of all images in that channel. The best-fit system parameters as determined by our pipeline are presented in Table \ref{doublefit_tab}. IRAC magnitudes in each channel for the primary stars are calculated from the best-fit primary flux, while for the confirmed or candidate low-mass companions the best-fit contrast is also used. The primary magnitudes span $7.30<[3.6]<12.89$ in Channel 1 and $5.35<[8.0]<12.70$ in Channel 4. The magnitudes of detected confirmed or candidate companions span $8.59<[3.6]<13.88$ in Channel 1 and $8.06<[8.0]<13.90$ in Channel 4. Three candidate low-mass companions were not detected by our pipeline (2MASS J05373850+2428517 c1, 2MASS J16111711--2217173 c1, and 2MASS J16151116--2420153 c1). The candidate companion to 2MASS J04554970+3019400 was detected in \textit{Gaia} DR2 \citep{gaia18} with measured $\pi=0.27\pm0.10$ mas, making it an unassociated background star. Photometry for the primary stars and candidate or confirmed companions of our target sample are shown in Table \ref{comb_spitzermag}.

In Table \ref{comp_mass} we derive companion masses for our target sample using our measured $[3.6]$ photometry (or $[4.5]$ for USco 1610--1913 which does not have Channel 1 images available) and the BT-Settl evolutionary models of \citet{allard12}. We use \textit{Gaia} DR2 distance estimates from \citet{bailer-jones18}, or if an estimate did not exist, the canonical distance to the star-forming region of which the system is a member to determine absolute magnitudes.

\begin{deluxetable*}{lccccccc}
\tablecaption{Best-Fit System Properties of Detected Companions and Neighbors}
\tablehead{2MASS & Other Name & Separation & Position Angle & $\Delta [3.6]$ & $\Delta [4.5]$ & $\Delta [5.8]$ & $\Delta [8.0]$\\
& & (arcsec) & (deg) & (mag) & (mag) & (mag) & (mag)}
\startdata
J03590986+2009361 B	& [SCH06] J0359099+2009362 B   	& $4.67\pm0.01$ 	& $264.2\pm0.1$	& $1.69\pm0.01$	& $1.56\pm0.01$	& $1.32\pm0.03$	& $0.95\pm0.06$	\\
J04233539+2503026 B	& FU Tau B               				& $5.68\pm0.04$ 	& $123.4\pm0.3$    	& $4.05\pm0.05$ 	& $4.32\pm0.04$ 	& $4.24\pm0.02$ 	& $4.12\pm0.01$ 	\\
J04292971+2616532 B	& FW Tau C					& $2.22\pm0.10$ 	& $292.6\pm2.2$    	& $4.74\pm0.08$ 	& $4.39\pm0.05$ 	& $4.43\pm0.07$ 	& $3.99\pm0.06$ 	\\
J04554970+3019400 c1\tablenotemark{a}	&				& $7.34\pm0.01$ 	& $129.0\pm0.2$	& $2.45\pm0.01$ 	& $2.63\pm0.01$ 	& $2.86\pm0.05$ 	& $3.33\pm0.12$ 	\\
J11075588--7727257 B 	& CHXR 28 B					& $1.87\pm0.04$ 	& $117.0\pm0.7$    	& $0.52\pm0.01$ 	& $0.47\pm0.01$ 	& $0.51\pm0.04$ 	& $0.60\pm0.03$ 	\\
J11122441--7637064 B	& Sz 41 B						& $1.97\pm0.01$ 	& $162.6\pm0.2$    	& $2.20\pm0.01$ 	& $2.06\pm0.01$ 	& $2.37\pm0.04$ 	& $2.72\pm0.02$ 	\\
J16101918--2502301 B	& USco 1610--2502 B			& $4.90\pm0.01$ 	& $241.2\pm0.3$    	& $2.63\pm0.02$ 	& $2.33\pm0.03$ 	& $2.18\pm0.01$ 	& $1.64\pm0.01$ 	\\
J16103196--1913062 B	& USco 1610--1913 B			& $5.83\pm0.01$ 	& $113.6\pm0.4$	& ...           		& $3.46\pm0.01$	& ...           		& $3.41\pm0.02$      
\enddata
\tablecomments{If an entry in $\Delta m$ is missing, no IRAC data existed for that object in that channel. The astrometric precision reflects that of the input priors. The candidate companions to 2MASS J05373850+2428517, 2MASS J16111711--2217173, and 2MASS J16151116--2420153 were not recovered.}
\tablenotetext{a}{The candidate companion to 2MASS J04554970+3019400 was detected in \textit{Gaia} DR2 \citep{gaia18} with measured $\pi=0.27\pm0.10$ mas, making it an unassociated background star.}
\label{doublefit_tab}
\end{deluxetable*}

\begin{turnpage}
\begin{deluxetable*}{lccccccccc}
\tablecaption{\textit{Spitzer}/IRAC Photometric Measurements for Test Case Sample}
\tablehead{2MASS & Other Name & $[3.6]_{\mathrm{P}}$ & $[4.5]_{\mathrm{P}}$ & $[5.8]_{\mathrm{P}}$ & $[8.0]_{\mathrm{P}}$ & $[3.6]_{\mathrm{N}}$ & $[4.5]_{\mathrm{N}}$ & $[5.8]_{\mathrm{N}}$ & $[8.0]_{\mathrm{N}}$ \\
& & (mag) & (mag) & (mag) & (mag) & (mag) & (mag) & (mag) & (mag)}
\startdata
J03590986+2009361	& [SCH06] J0359099+2009362		& $12.20\pm0.02$	& $12.08\pm0.02$	& $12.05\pm0.02$	& $12.00\pm0.03$	& $13.88\pm0.02$	& $13.64\pm0.02$	& $13.37\pm0.04$	& $12.95\pm0.06$	\\
J04233539+2503026	& FU Tau						& $8.35\pm0.02$	&  $7.74\pm0.02$	& $7.29\pm0.02$    	& $6.53\pm0.02$	& $12.39\pm0.05$    	& $12.06\pm0.05$    	& $11.53\pm0.02$    	& $10.65\pm0.02$ 	\\
J04292971+2616532	& FW Tau						& $9.01\pm0.02$	&  $8.90\pm0.02$	& $8.84\pm0.02$    	& $8.85\pm0.02$	& $13.75\pm0.08$    	& $13.29\pm0.05$    	& $13.27\pm0.07$    	& $12.84\pm0.06$ 	\\
J04554970+3019400	&							& $11.34\pm0.02$	& $11.15\pm0.02$	& $10.93\pm0.02$	& $10.58\pm0.02$ 	& $13.80\pm0.02$    	& $13.78\pm0.02$    	& $13.79\pm0.05$    	& $13.91\pm0.12$ 	\\
J05373850+2428517	& [SCH06] J0537385+2428518		& $10.41\pm0.02$	& $10.33\pm0.02$	& $10.25\pm0.02$    	& $10.23\pm0.02$	& $>12.37$		& $>13.25$		& $>13.65$		& $>13.82$		\\
J11075588--7727257	& CHXR 28	            			& $7.99\pm0.02$	&  $8.00\pm0.02$	& $7.81\pm0.02$   	& $7.83\pm0.02$	& $8.51\pm0.02$    	& $8.47\pm0.02$    	& $8.32\pm0.04$    	&  $8.44\pm0.04$ 	\\
J11122441--7637064	& Sz 41                      			& $7.31\pm0.02$	&  $6.87\pm0.02$	& $6.39\pm0.02$    	& $5.35\pm0.02$	& $9.50\pm0.02$    	& $8.93\pm0.02$    	& $8.76\pm0.05$    	&  $8.07\pm0.03$ 	\\
J16101918--2502301	& USco 1610--2502				& $8.27\pm0.02$	&  $8.33\pm0.02$	& $8.25\pm0.02$    	& $8.19\pm0.02$	& $10.90\pm0.03$    	& $10.67\pm0.03$    	& $10.43\pm0.02$    	&  $9.83\pm0.02$ 	\\
J16103196--1913062	& USco 1610--1913				& ... 				&  $8.68\pm0.02$	& ...               		& $8.58\pm0.02$	& ...               		& $12.13\pm0.02$    	& ...               		& $12.00\pm0.02$ 	\\
J16111711--2217173		& [SCH06] J16111711--22171749	& $12.86\pm0.02$    	& $12.75\pm0.02$    	& $12.66\pm0.03$    	& $12.72\pm0.04$	& $>17.75$		& $>17.66$		& $>15.40$		& $>14.81$		\\ 
J16151116--2420153	& [SCH06] J16151115--24201556	& $12.79\pm0.02$    	& $12.71\pm0.02$	& $12.64\pm0.02$    	& $12.69\pm0.04$	& $>17.79$		& $>17.71$		& $>15.68$		& $>14.56$
\enddata
\tablecomments{ Subscript ``P" denotes the primary star of the test case sample system while subscript ``N" denotes the confirmed companion, candidate companion, or neighbor. If an entry is missing, no IRAC data existed for that object in that channel.}
\label{comb_spitzermag}
\end{deluxetable*}
\end{turnpage}

\begin{figure}
\centering
\includegraphics[trim={1.5cm 0 0 0},width=0.47\textwidth]{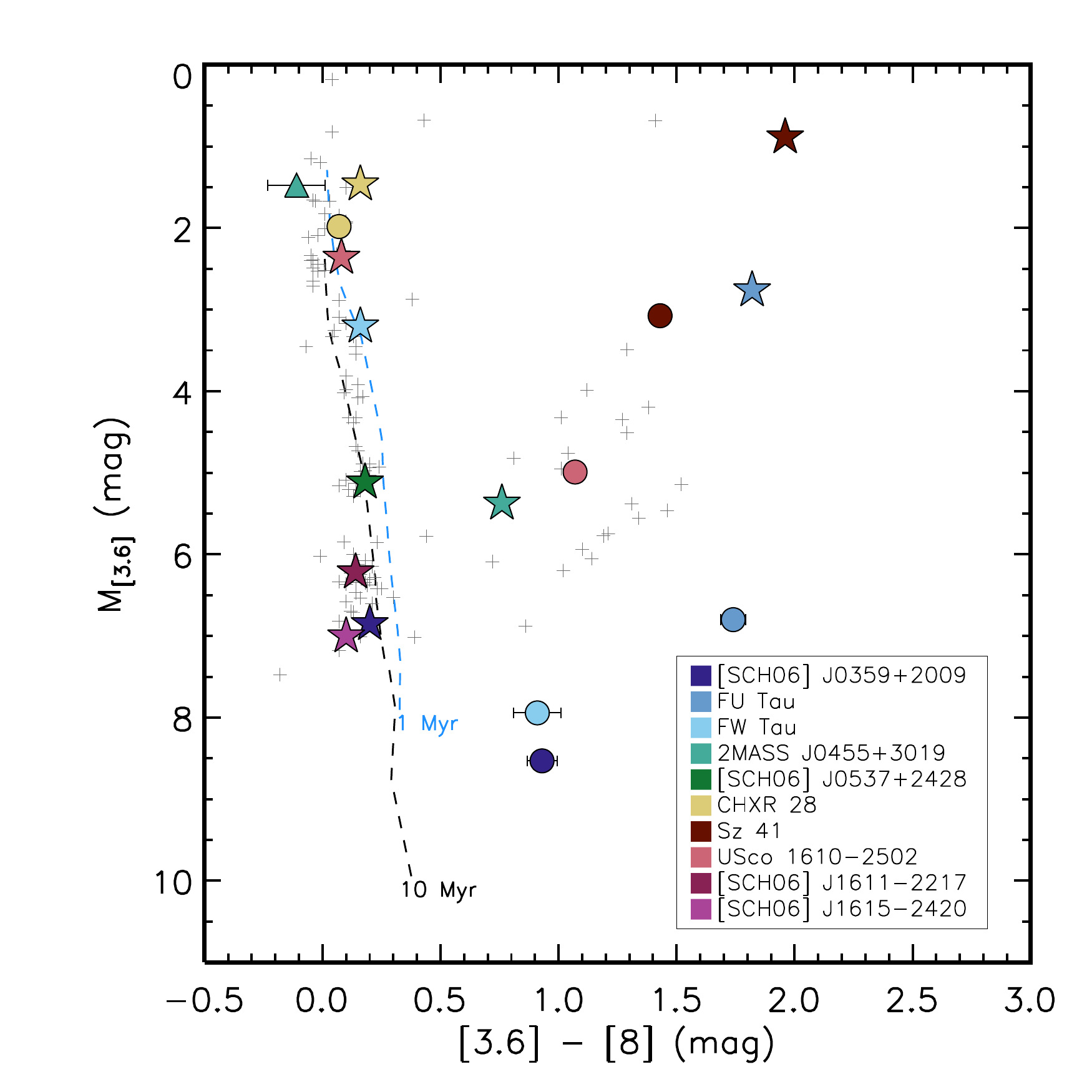}
\caption{\textit{Spitzer}/IRAC color-magnitude diagram for our target systems and Upper Scorpius members with IRAC [3.6] and [8.0] measurements from \citet{luhman12}. Absolute $[3.6]$ magnitudes were determined using \textit{Gaia} DR2 distance estimates from \citet{bailer-jones18}, or if an estimate did not exist, the canonical distance to the star-forming region of which the system is a member. The primary components of the target sample members are indicated as stars while the confirmed companions are indicated as filled circles. The nearby neighbor to 2MASS J0455+3019 is an unassociated background star and is denoted by a teal triangle. Also indicated are the intrinsic photospheric $[3.6]-[8.0]$ colors from BT-Settl models of \citet{allard12} at 1 and 10 Myr, using dashed light blue and black lines, respectively. Three sample primaries and five secondaries have $[3.6]-[8.0]$ colors indicative of a circumstellar or circum(sub)stellar disks.}
\label{bts_iso_cmd}
\end{figure}

In Figure \ref{bts_iso_cmd} we present a color-magnitude diagram of $[3.6]-[8.0]$ versus $[3.6]$ showing the 10 system primaries, four confirmed low-mass companions, and one candidate companion detected in Channels 1 and 4, as well as Upper Scorpius members with both \textit{Gaia} DR2 parallaxes and \textit{Spitzer}/IRAC photometric measurements from \cite{luhman12}. We also show the intrinsic photospheric mid-infrared color-magnitude sequences from BT-Settl models \citep{allard12} for 1 and 10 Myr objects in dashed light blue and black lines, respectively. We use these $[3.6]-[8.0]$ colors to assess the presence of excess emission due to a disk since intrinsic photospheric colors can vary with spectral type. Three primaries have $[3.6]-[8.0]$ color more than 3-$\sigma$ above their intrinsic photosphere color (Sz 41, FU Tau, and 2MASS J0455+3019) while one candidate and four confirmed wide companions have $[3.6]-[8.0]$ color more than 3-$\sigma$ above their expected intrinsic photosphere color. Tables \ref{prim_colors} and \ref{comp_colors} list the primary and secondary IRAC colors of our target sample across all channels, respectively.

\begin{deluxetable*}{lcccccc}
\tablecaption{Derived Masses}
\tablehead{2MASS		&	Other Name				&	Distance			&	$M_{[3.6]}$		&	Age		& Reference	& Mass\tablenotemark{a}			\\
					&							&	(pc)				&	(mag)			&	(Myr)		&			& }
\startdata
J03590986+2009361 B	& [SCH06] J0359099+2009362 B	&	$117.4\pm2.3$		&	$8.53\pm0.05$		&	$5-10$	&	1		& $15-16$	 $M_{\mathrm{Jup}}$	\\
J04233539+2503026 B	& FU Tau B					&	$131.2\pm2.6$		&	$6.80\pm0.07$		&	$1-2$	&	2		& $22-29$	 $M_{\mathrm{Jup}}$	\\
J04292971+2616532 C	& FW Tau C					&	$145\pm15$		&	$7.94\pm0.24$		&	$1-2$	& 	2		& $10-14$ $M_{\mathrm{Jup}}$	\tablenotemark{b}\\
J04554970+3019400 c1 	&							&	$2920\pm1360$	&	$0.96\pm0.80$		& ...			& ...			& ...	\\
J05373850+2428517 c1	& [SCH06] J0537385+2428518 c1	&	$114.5\pm1.2$		&	$>7.08$			&	$1-2$	&	1		& $<25$ $M_{\mathrm{Jup}}$	\\
J11075588--7727257 B	& CHXR 28 B					&	$202.1\pm10.4$	&	$1.98\pm0.11$		&	$2-3$	&	3		& $1.2-1.5$ $M_{\odot}$	\\
J11122441--7637064 B	& Sz 41 B						&	$192.7\pm0.8$		&	$3.08\pm0.02$		&	$2-3$	&	3		& $0.47-0.61$ $M_{\odot}$	\\
J16101918--2502301 B	& USco 1610--2502 B			&	$152.0\pm1.9$		&	$4.99\pm0.04$		&	$5-10$	&	4, 5		& $0.18-0.28$ $M_{\odot}$	\\
J16103196--1913062 B	& USco 1610--1913 B			&	$132.9\pm1.3$		&	$6.51\pm0.03$\tablenotemark{c}	&	$5-10$	& 4, 5	& $34-73$	 $M_{\mathrm{Jup}}$	\\
J16111711--2217173	 c1	& [SCH06] J16111711--22171749 c1	&	$213.1\pm18.3$	&	$>11.11$			&	$5-10$	&	4, 5		& $<10$ $M_{\mathrm{Jup}}$		\\
J16151116--2420153 c1	& [SCH06] J16151115--24201556 c1	&	$143.8\pm4.5$		&	$>12.00$			&	$5-10$	&	4, 5		& $<10$ $M_{\mathrm{Jup}}$
\enddata
\tablecomments{Age ranges for target systems obtained from the following references: (1) \citet{slesnick06}; (2) \citet{kraus09a}; (3) \citet{luhman04}; (4) \citet{deZeeuw99}; (5) \citet{pecaut12}.}
\tablenotetext{a}{A model-derived mass is not meaningful for the neighbor to J04554970+3019400 because it is a background star. For candidate companions that could not be confirmed, we list the mass or mass limit that would be consistent with that photometry.}
\tablenotetext{b}{The nature of FW Tau C is a matter of debate since different aspects of its observations are consistent with either a substellar companion surrounded by an edge-on disk or a PMC embedded in a low inclination disk. See Sections \ref{fwtau1} and \ref{fwtau2} for further discussion.}
\tablenotetext{c}{No [3.6] images were available for USco 1610--1913. The listed absolute magnitude and derived mass are based on the [4.5] magnitude measured in this work.}
\label{comp_mass}
\end{deluxetable*}

\begin{turnpage}
\begin{deluxetable*}{lcccccccc}
\tablecaption{Measured Mid-IR Colors for Primaries}
\tablehead{2MASS 		&	Other Name				&	$[3.6]-[4.5]$		&	$[3.6]-[5.8]$		&	$[3.6]-[8.0]$		&	$[4.5]-[5.8]$		&	$[4.5]-[8.0]$		&	$[5.8]-[8.0]$		&	Disk?\\
					&							&	(mag)			&	(mag)			&	(mag)			&	(mag)			&	(mag)			&	(mag)			&	Y/N	}
\startdata
J03590986+2009361	& [SCH06] J0359099+2009362		&	$0.11\pm0.02$		&	$0.15\pm0.03$		&	$0.20\pm0.03$		&	$0.04\pm0.03$		&	$0.09\pm0.03$		&	$0.05\pm0.03$		&	N	\\
J04233539+2503026	& FU Tau A					&	$0.61\pm0.02$		&	$1.06\pm0.02$		&	$1.82\pm0.02$		&	$0.45\pm0.02$		&	$1.21\pm0.02$		&	$0.76\pm0.02$		&	Y	\\
J04292971+2616532	& FW Tau	AB					&	$0.11\pm0.02$		&	$0.17\pm0.02$		&	$0.16\pm0.02$		&	$0.06\pm0.02$		&	$0.05\pm0.02$		&	$-0.01\pm0.02$		&	N	\\
J04554970+3019400 	&							&	$0.19\pm0.02$		&	$0.42\pm0.02$		&	$0.77\pm0.02$		&	$0.23\pm0.02$		&	$0.58\pm0.02$		&	$0.35\pm0.02$		&	Y	\\
J05373850+2428517	& [SCH06] J0537385+2428518		&	$0.09\pm0.02$		&	$0.16\pm0.02$		&	$0.18\pm0.02$		&	$0.08\pm0.02$		&	$0.09\pm0.02$		&	$0.02\pm0.02$		&	N	\\
J11075588--7727257	& CHXR 28 Aa,Ab 				&	$-0.01\pm0.02$		&	$0.18\pm0.03$		&	$0.16\pm0.03$		&	$0.18\pm0.03$		&	$0.16\pm0.03$		&	$-0.02\pm0.03$		&	N	\\
J11122441--7637064	& Sz 41 A						&	$0.44\pm0.02$		&	$0.92\pm0.02$		&	$1.96\pm0.02$		&	$0.48\pm0.02$		&	$1.53\pm0.02$		&	$1.04\pm0.02$		&	Y	\\
J16101918--2502301	& USco 1610--2502 A			&	$-0.06\pm0.02$		&	$0.02\pm0.02$		&	$0.08\pm0.02$		&	$0.08\pm0.02$		&	$0.14\pm0.02$		&	$0.06\pm0.02$		&	N	\\
J16103196--1913062	& USco 1610--1913 A			&	...				&	...				&	...				&	...				&	$0.09\pm0.02$		&	...				&	N	\\
J16111711--2217173		& [SCH06] J16111711--22171749	&	$0.11\pm0.02$		&	$0.20\pm0.03$		&	$0.14\pm0.04$		&	$0.09\pm0.03$		&	$0.03\pm0.04$		&	$-0.06\pm0.05$		&	N	\\
J16151116--2420153	& [SCH06] J16151115--24201556	&	$0.07\pm0.02$		&	$0.15\pm0.03$		&	$0.10\pm0.04$		&	$0.07\pm0.03$		&	$0.02\pm0.04$		&	$-0.05\pm0.04$		&	N
\enddata
\tablecomments{If an entry is missing, either no IRAC data existed for that object or no images were adequately fit in that IRAC Channel.}
\label{prim_colors}
\end{deluxetable*}
%%%
\begin{deluxetable*}{lccccccccc}
\tablecaption{Measured Mid-IR Colors for Detected Companions and Neighbors}
\tablehead{2MASS		&	Other Name				&	$[3.6]-[4.5]$		&	$[3.6]-[5.8]$		&	$[3.6]-[8.0]$		&	$[4.5]-[5.8]$		&	$[4.5]-[8.0]$		&	$[5.8]-[8.0]$		&	Disk?\\
					&							&	(mag)			&	(mag)			&	(mag)			&	(mag)			&	(mag)			&	(mag)			&	Y/N}
\startdata
J03590986+2009361 B	& [SCH06] J0359099+2009362	B	&	$0.24\pm0.03$		&	$0.51\pm0.04$		&	$0.94\pm0.07$		&	$0.27\pm0.04$		&	$0.69\pm0.07$		&	$0.42\pm0.07$		&	Y	\\
J04233539+2503026 B	& FU Tau B					&	$0.33\pm0.07$		&	$0.86\pm0.06$		&	$1.74\pm0.05$		&	$0.53\pm0.05$		&	$1.41\pm0.05$		&	$0.88\pm0.03$		&	Y	\\
J04292971+2616532 C	& FW Tau	C					&	$0.46\pm0.10$		&	$0.48\pm0.11$		&	$0.91\pm0.10$		&	$0.02\pm0.09$		&	$0.45\pm0.08$		&	$0.44\pm0.09$		&	Y	\\
J04554970+3019400 c1 	&							&	$0.02\pm0.03$		&	$0.00\pm0.05$		&	$-0.11\pm0.12$		&	$-0.01\pm0.05$		&	$-0.13\pm0.12$		&	$-0.12\pm0.13$		&	N	\\
J11075588--7727257 B	& CHXR 28 B					&	$0.05\pm0.03$		&	$0.19\pm0.05$		&	$0.08\pm0.04$		&	$0.15\pm0.05$		&	$0.03\pm0.04$		&	$-0.12\pm0.06$		&	N	\\
J11122441--7637064 B	& Sz 41 B						&	$0.57\pm0.03$		&	$0.75\pm0.05$		&	$1.44\pm0.03$		&	$0.18\pm0.05$		&	$0.86\pm0.03$		&	$0.69\pm0.05$		&	Y	\\
J16101918--2502301 B	& USco 1610--2502 B			&	$0.24\pm0.04$		&	$0.47\pm0.03$		&	$1.08\pm0.03$		&	$0.23\pm0.04$		&	$0.84\pm0.03$		&	$0.61\pm0.02$		&	Y	\\
J16103196--1913062 B	& USco 1610--1913 B			&	...				&	...				&	...				&	...				&	$0.14\pm0.03$		&	...				&	N
\enddata
\tablecomments{If an entry is missing, either no IRAC data existed for that object or no images were adequately fit in that IRAC Channel.}
\label{comp_colors}
\end{deluxetable*}
\end{turnpage}

\begin{figure}
\centering
\includegraphics[trim={1.5cm 0 0 0},width=0.47\textwidth]{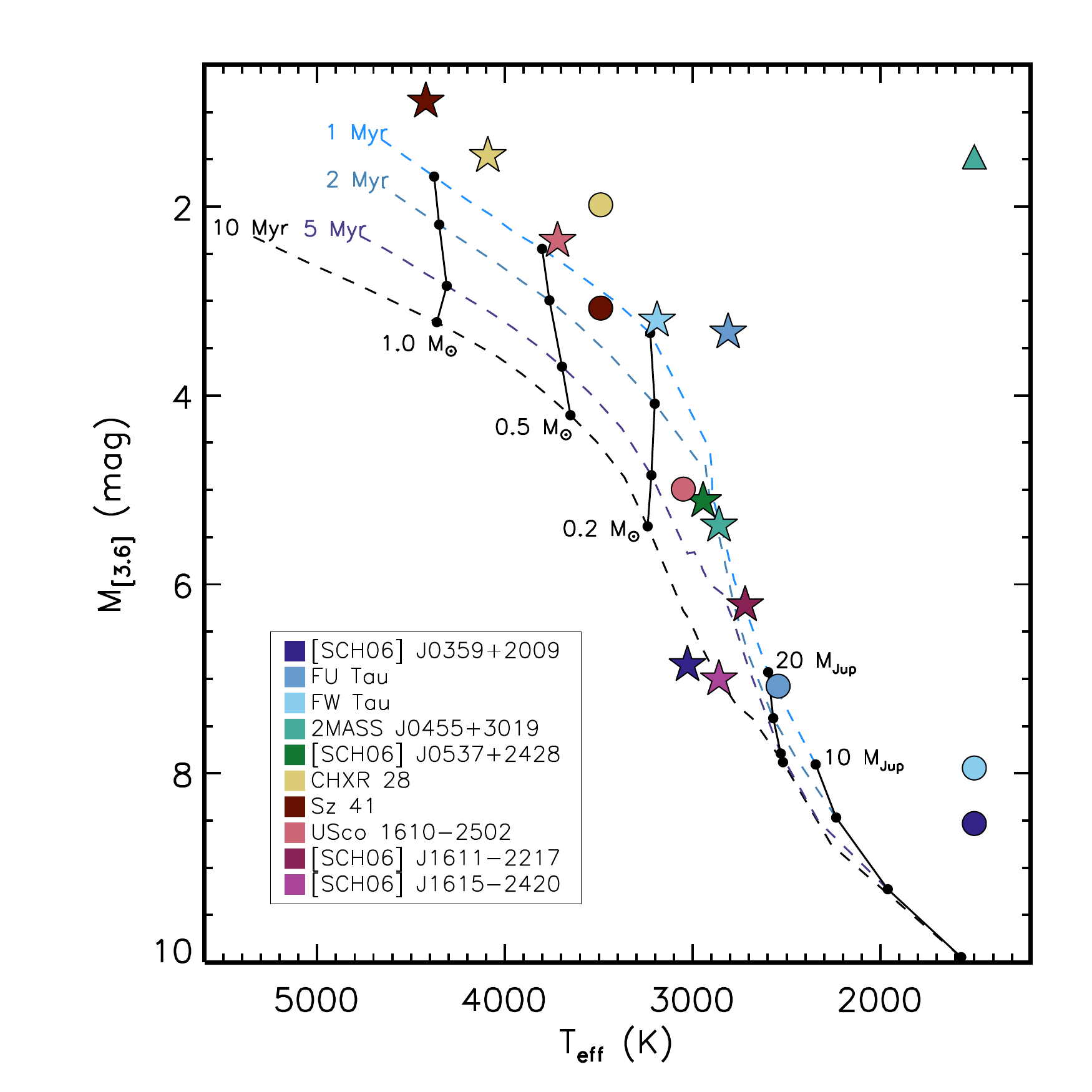}
\caption{H-R diagram for our target systems. The primary components of the target sample members are indicated as stars and confirmed secondary components indicated as circles. We use $T_{\mathrm{eff}}$ measurements reported in the literature for objects that were spectroscopically derived from spectral types, or we estimate them based on optical spectral types from the literature and the spectral type to temperature conversion from \citet{herczeg14}. The 1, 2, 5, and 10 Myr isochrones of \citet{allard12} are plotted (dashed lines) with 10 $M_{\mathrm{Jup}}$, 20 $M_{\mathrm{Jup}}$, 0.2 $M_\odot$, 0.5 $M_\odot$, and 1.0 $M_\odot$ iso-mass tracks (solid lines). The secondary components to [SCH06] J0359+2009 (dark blue circle) and FW Tau (light blue circle), and nearby neighbor to 2MASS J0455+3019 (teal triangle) are indicated at their corresponding $M_{[3.6]}$ but not placed on this diagram because they do not have independent temperature estimates.}
\label{bts_iso_hrd}
\end{figure}

\subsection{Notes on Individual Systems}
Our reprocessing of IRAC images yielded detections of one candidate and six confirmed low-mass companions and one unassociated neighbor. Five of the companions have $[3.6]-[8.0]$ colors more than 3-$\sigma$ above their expected intrinsic photosphere color. We describe four of them in more detail in the following sections.

\subsubsection{[SCH06] J0359+2009 B: A New Wide Companion Near the Planet--Brown Dwarf Boundary}
\label{schj0359}
\citet{slesnick06} first identified 2MASS J03590986+2009361 (hereafter [SCH06] J0359+2009) as a potential member of an older distributed population of Taurus, proposed by \citet{wichmann96}. Observational follow-up conducted by \citet{kraus12} identified a candidate low-mass companion in the vicinity of [SCH06] J0359+2009 at projected separation $\rho=4\farcs66$. No further targeted observations to confirm association have been reported in the literature and the nature of the candidate companion has remained unclear.

\begin{figure*}
\centering
\includegraphics[trim={0.45in 0.15in 0.35in 0.15in},clip,width=0.75\textwidth]{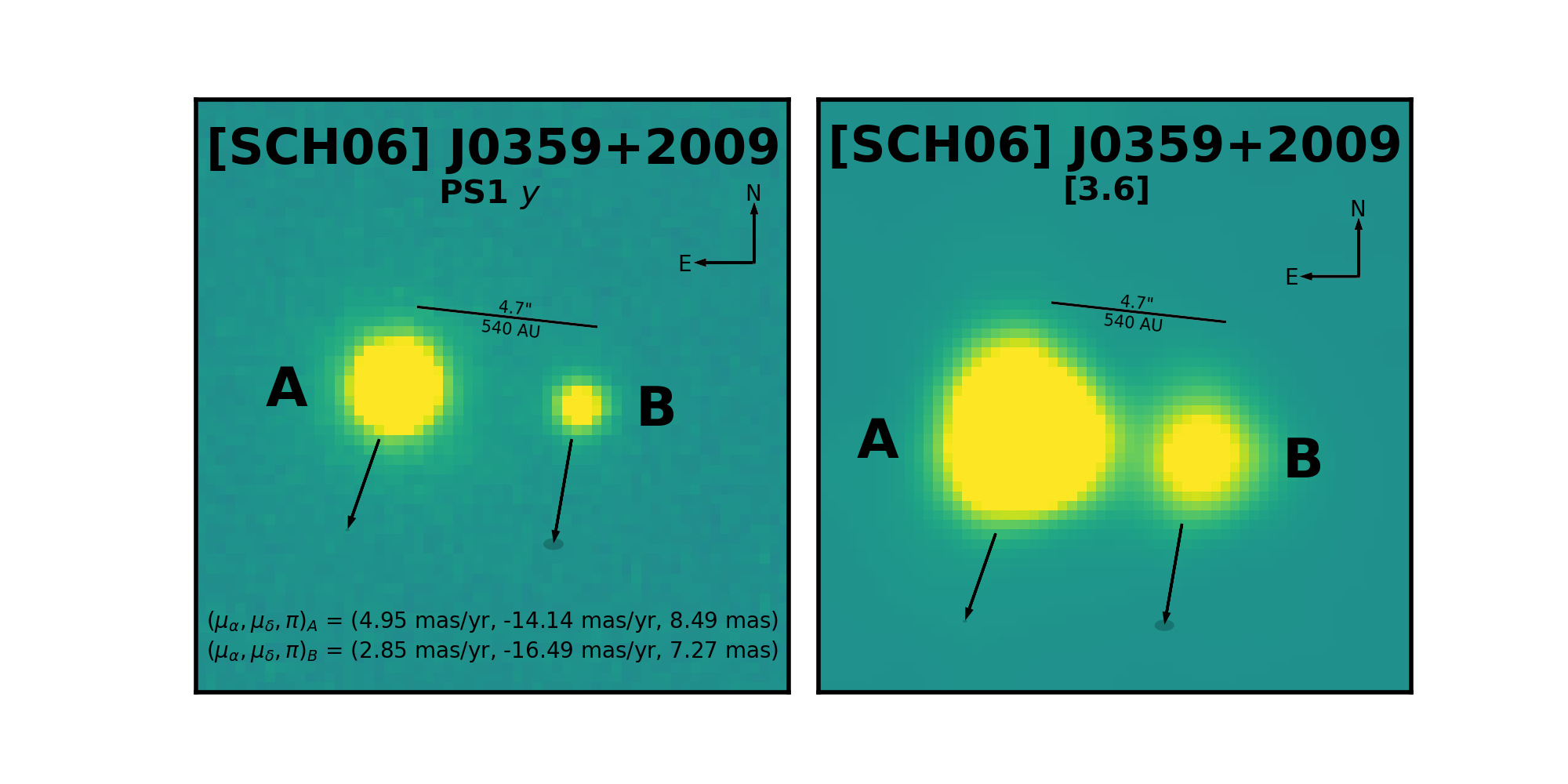}
\caption{Pan-STARRS $y$-band (left) and IRAC Ch 1 (right) images of the [SCH06] J0359+2009 system. [SCH06] J0359+2009 B is located at a separation of $4\farcs7$ (540 au). Proper motion vectors for [SCH06] J0359+2009 A and B are shown as black arrows indicating the direction of motion; the length of the vectors have been magnified to scale to make them visible. The proper motion errors are also shown at the ends of the proper motion vectors as shaded gray ellipses. Given the component masses and projected separation (see Section \ref{schj0359}), this system appears to be an older analog of ultrawide brown dwarf pairs like FU Tau \citep{luhman09}.}
\label{schj0359_discovery}
\end{figure*}

\textit{Gaia} DR2 \citep{gaia18} measure $(\mu_\alpha,\mu_\delta,\pi)=(4.95\pm0.37$ mas $\mathrm{yr^{-1}}, -14.14\pm0.20$ mas $\mathrm{yr^{-1}}, 8.49\pm0.16$ mas) for the primary and $(\mu_\alpha,\mu_\delta,\pi)=(2.85\pm1.69$ mas $\mathrm{yr^{-1}}, -16.49\pm0.98$ mas $\mathrm{yr^{-1}}, 7.27\pm0.87$ mas) for the candidate companion. These measurements are consistent with the expected Taurus proper motion ($\mu=(+6,-20)$ mas $\mathrm{yr^{-1}}$; \citealp{kraus17}), as well as comovement and codistance for the two objects to within 1.5-$\sigma$. With these additional data we confirm association and report the discovery of a new low-mass companion (hereafter [SCH06] J0359+2009 B; see Figure \ref{schj0359_discovery}).

\begin{figure}
\centering
\includegraphics[trim={0 0 0 0},width=0.35\textwidth,angle=90]{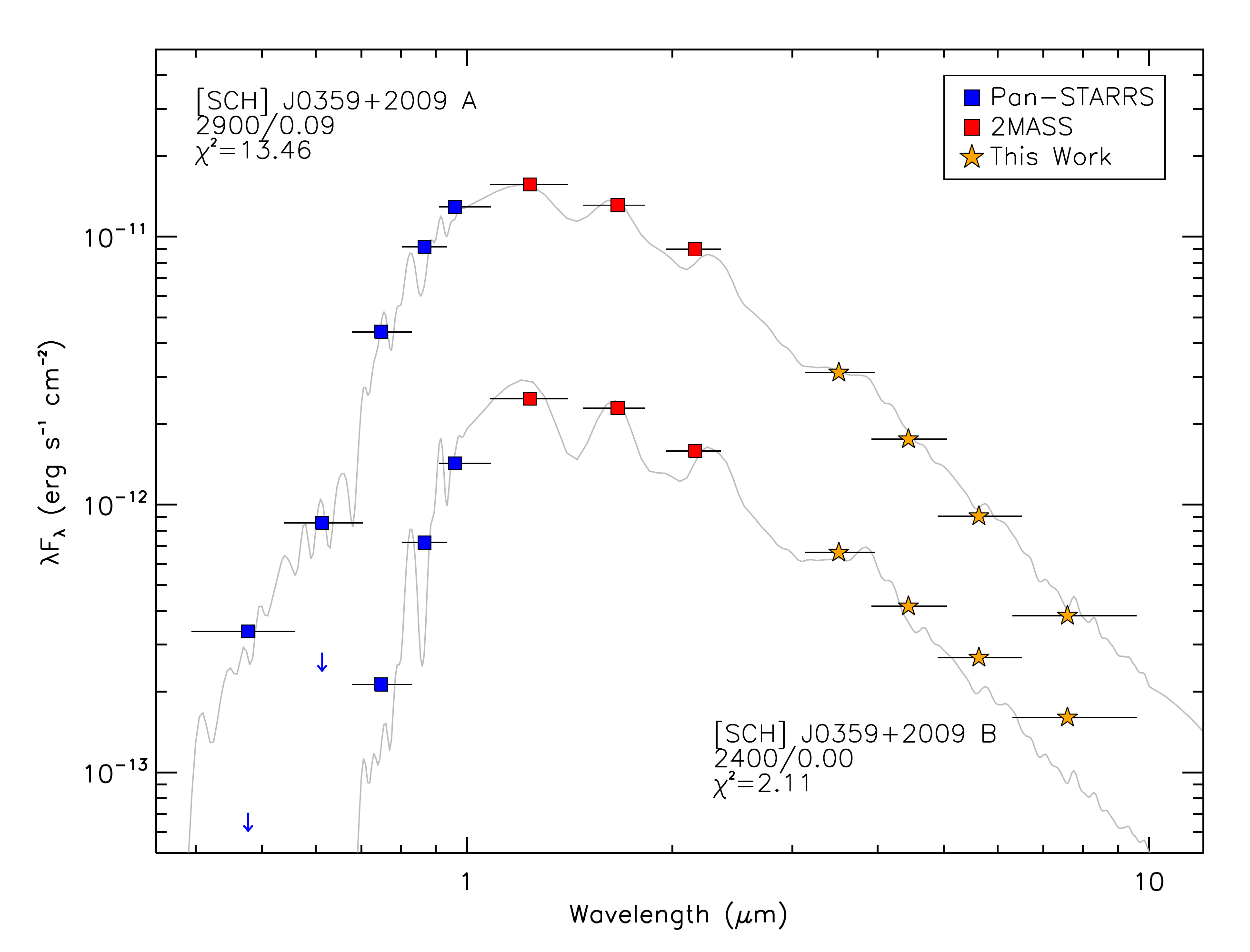}
\caption{Spectral energy distributions of [SCH06] J0359+2009 and its companion [SCH06] J0359+2009 B. Photometric uncertainties are smaller than the symbol sizes. An $E(B-V)=0.09^{+0.02}_{-0.015}$ mag, $T_\mathrm{eff}=2900\pm50$ K BT-Settl model best fits the photometry of the primary for wavelengths $>0.4$ $\mu$m. An $E(B-V)=0.00\pm0.03$ mag, $T_\mathrm{eff}=2400\pm50$ K BT-Settl model best fits the $0.7-3.6$ $\mu$m photometry of the companion. Both models are plotted in gray. The bandpasses of the photometric filters are plotted as horizontal lines. The excess flux seen in $[5.8]$ and $[8.0]$ is likely due to the presence of a circumplanetary/circum(sub)stellar disk.}
\label{sed_schj0359}
\end{figure}

In Table \ref{schj0359_phot} we summarize Pan-STARRS optical (PS1; \citealp{chambers16}), 2MASS near-infrared \citep{cutri03} and the \textit{Spitzer}/IRAC mid-infrared photometry measured in this work for both [SCH06] J0359+2009 A and [SCH06] J0359+2009 B. We use these data to analyze the SEDs of the [SCH06] J0359+2009 system. We fit solar metallicity BT-Settl model atmospheres \citep{allard12} spanning effective temperatures between 2000 and 3500 K ($\Delta T_{\mathrm{eff}}=100$ K) fixed at $\log$ g $= 4.0$, which is appropriate for late M dwarfs with ages $\tau>5$ Myr from evolutionary models. We also fit for $E(B-V)$ using the extinction curve of \citet{fitzpatrick99} spanning from 0.0 to 0.2 mag in steps of 0.01 mag. For the primary we fit PS1 $grizy$, 2MASS $JHK$ and \textit{Spitzer}/IRAC photometric data using $\chi^2$ minimization. We fit only the PS1 $rizy$, 2MASS $JHK$ and IRAC $[3.6]$ photometric data for the companion. We present the SEDs of both [SCH06] J0359+2009 and [SCH06] J0359+2009 B in Figure \ref{sed_schj0359}.

\begin{figure}
\centering
\includegraphics[trim={0 0 0 0},width=0.35\textwidth,angle=90]{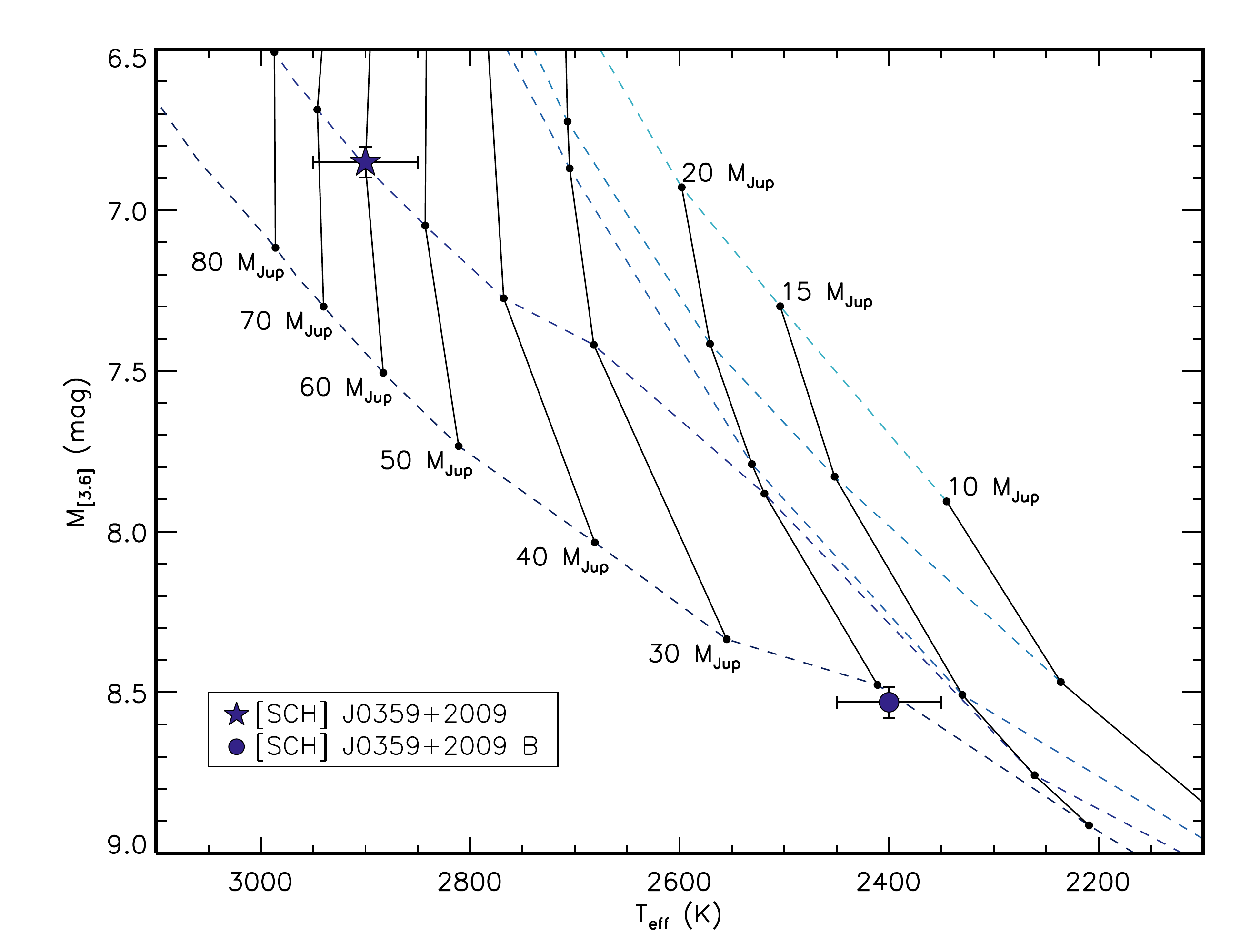}
\caption{H-R diagram for [SCH06] J0359+2009 and its companion. For both components, the temperatures are estimated from our SED fits (Section 5.2.1; Figure \ref{sed_schj0359}). The 1, 2, 5, 10, and 20 Myr isochrones of \citet{allard12} are plotted (dashed lines) with mass tracks (solid lines) from 10 $M_{\mathrm{Jup}}$ to 80 $M_{\mathrm{Jup}}$. The H-R diagram positions are nominally consistent with isochronal ages of 10-20 Myr, though those ages appear to be underestimated by most current models (e.g., \citealt{feiden16}). The position of the primary indicates a mass of $60 \pm 10$ $M_{\mathrm{Jup}}$ while the position of the companion indicates a mass of $20 \pm 5$ $M_{\mathrm{Jup}}$.}
\label{hrd_schj0359}
\end{figure}

The best-fitting model for [SCH06] J0359+2009 A is $T_{\mathrm{eff}}=2900\pm50$ K and $E(B-V)=0.09^{+0.02}_{-0.015}$ mag while for [SCH06] J0359+2009 B the best-fitting model is $T_{\mathrm{eff}}=2400\pm50$ K and $E(B-V)=0.00\pm0.03$ mag. Although the best-fit SEDs of [SCH06] J0359+2009 A and B have discrepant $E(B-V)$ of $\sim0.1$ mag, wide binary pairs in Taurus have been found to have differing reddening values of similar amounts \citep{herczeg14} due to the systematic uncertainty in atmospheric models of young low-mass objects (e.g., \citealt{dupuy10}), in addition to the young objects themselves likely harboring spots that change the emergent spectrum (e.g., \citealt{gully17}). We infer system masses from predictions of BT-Settl evolutionary models (see Figure \ref{hrd_schj0359}) based on their absolute $[3.6]$ magnitude and the best-fitting model $T_{\mathrm{eff}}$. We estimate [SCH06] J0359+2009 A to have a mass of $60 \pm 10 M_{\mathrm{Jup}}$ and the companion to have a mass of $20 \pm 5 M_{\mathrm{Jup}}$. The H-R diagram positions are nominally consistent with isochronal ages of 10-20 Myr, though those ages appear to be underestimated by most current models (e.g., \citealt{feiden16}). As seen in Figure \ref{sed_schj0359}, the \textit{Spitzer}/IRAC 8 $\mu$m photometry for [SCH06] J0359+2009 B disagrees with the best-fitting model at $>$8-$\sigma$. This is consistent with our measured mid-infrared excess of $[3.6]-[8.0]=0.94\pm0.07$ mag which is discrepant with the color of a M9 photosphere at the 7.9-$\sigma$ level and L0 photosphere at the 4.4-$\sigma$ level as measured empirically by \citet{luhman10}. Given the component masses and projected separation, this system appears to be an older analog of ultrawide brown dwarf pairs like FU Tau \citep{luhman09}.

\citet{kraus12} identified a second candidate companion in the vicinity of [SCH06] J0359+2009 A at projected separation $\rho=5\farcs95$ and P.A.$=99^\circ$. This source was not in our original target sample nor do we detect it (see Figure \ref{stacked_residuals}, top row). We derive upper limits on its IRAC photometry of $[3.6]>18.8$ mag, $[4.5]>18.0$ mag, $[5.8]>16.5$ mag, and $[8.0]>15.4$ mag based on our detection limits discussed in Section \ref{dls}.

\begin{deluxetable}{lcccc}
\tablecaption{[SCH06] J0359+2009 Photometry}
\tablehead{Filter & Wavelength & A Flux & B Flux & Reference \\
& ($\mu$m) & (mag) & (mag) & }
\startdata
$g_{\mathrm{P1}}$	& 0.481	& $19.56\pm0.02$\tablenotemark{a}	& $>21.25$\tablenotemark{a}	& 1 \\
$r_{\mathrm{P1}}$	& 0.617	& $18.28\pm0.02$\tablenotemark{a}	& $>19.49$\tablenotemark{a}		& 1 \\
$i_{\mathrm{P1}}$	& 0.752	& $16.29\pm0.02$\tablenotemark{a}	& $19.58\pm0.02$\tablenotemark{a}	& 1 \\
$z_{\mathrm{P1}}$	& 0.866	& $15.34\pm0.02$\tablenotemark{a}	& $18.10\pm0.02$\tablenotemark{a}	& 1 \\
$y_{\mathrm{P1}}$	& 0.962	& $14.86\pm0.02$\tablenotemark{a}	& $17.25\pm0.02$\tablenotemark{a}	& 1 \\
$J$				& 1.235	& $13.48\pm0.03$	& $15.48\pm0.07$	& 3 \\	
$H$				& 1.662	& $12.83\pm0.03$	& $14.79\pm0.06$	& 3 \\	
$K_s$			& 2.159	& $12.53\pm0.03$	& $14.42\pm0.07$	& 3 \\
$[3.6]$			& 3.6		& $12.20\pm0.02$	& $13.88\pm0.02$	& This work \\
$[4.5]$			& 4.5		& $12.08\pm0.02$	& $13.64\pm0.02$	& This work \\
$[5.8]$			& 5.8		& $12.05\pm0.02$	& $13.37\pm0.04$	& This work \\
$[8.0]$			& 8.0		& $12.00\pm0.03$	& $12.95\pm0.06$	& This work 
\enddata
\tablenotetext{a}{AB magnitudes}
\tablerefs{(1) \citet{chambers16}, (2) \citet{lawrence12}, (3) \citet{cutri03}}
\label{schj0359_phot}
\end{deluxetable}

\subsubsection{FU Tau}
FU Tau AB is an isolated wide binary with $\rho=5\farcs7$ \citep{luhman09}, or 750 au at its parallactic distance (131.2 pc; \citealt{gaia18}). Using spectroscopic observations in the optical, \citet{luhman09} estimate the spectral type of the primary to be M7.25 and the secondary to be M9.25, deriving model-dependent masses for FU Tau A and B to be 50 $M_{\mathrm{Jup}}$ and 15 $M_{\mathrm{Jup}}$, respectively. We clearly detect FU Tau B in all four IRAC channels. Both FU Tau A and B show significant excess with $[3.6]-[8.0]=1.83\pm0.02$ mag for the primary and $[3.6]-[8.0]=1.77\pm0.08$ mag for the secondary. This confirms the previous detection of this disk as detailed in \citet{luhman09}. FU Tau A also shows signs of variability. Observations taken in 2007 exist of FU Tau in all four channels. Observations were also taken for FU Tau in 2005, but only in Channels 2 and 4. Visual inspection of the residual images produced by our pipeline suggested Channels 2 and 4 were not well fit by the same contrast values across both epochs. We ran the individual epochs of FU Tau observations through our pipeline to quantify the variability. We measure $[4.5]=7.67\pm0.02$ and $[8.0]=6.41\pm0.02$ for the 2005 observations while we measure $[4.5]=7.84\pm0.02$ and $[8.0]=6.68\pm0.02$ in the 2007 images. The FU Tau B photometry does not change significantly between epochs, suggesting the variability is due to the primary. In addition to this, FU Tau A is overluminous compared to the 1 Myr isochrone. \citet{luhman09} posit that this may be due to FU Tau A itself being an unresolved binary, though this solution would not account for all of the overluminosity. \citet{stelzer10} obtained \textit{Chandra} X-ray observations which suggested that atypical magnetic activity or accretion is present in FU Tau A and hence that severe rotation and magnetic field effects might be reducing the efficiency of convection, affecting its place on the H--R diagram.

\begin{figure}
\centering
\includegraphics[trim={0 3cm 0 5cm},width=0.27\textwidth,angle=90]{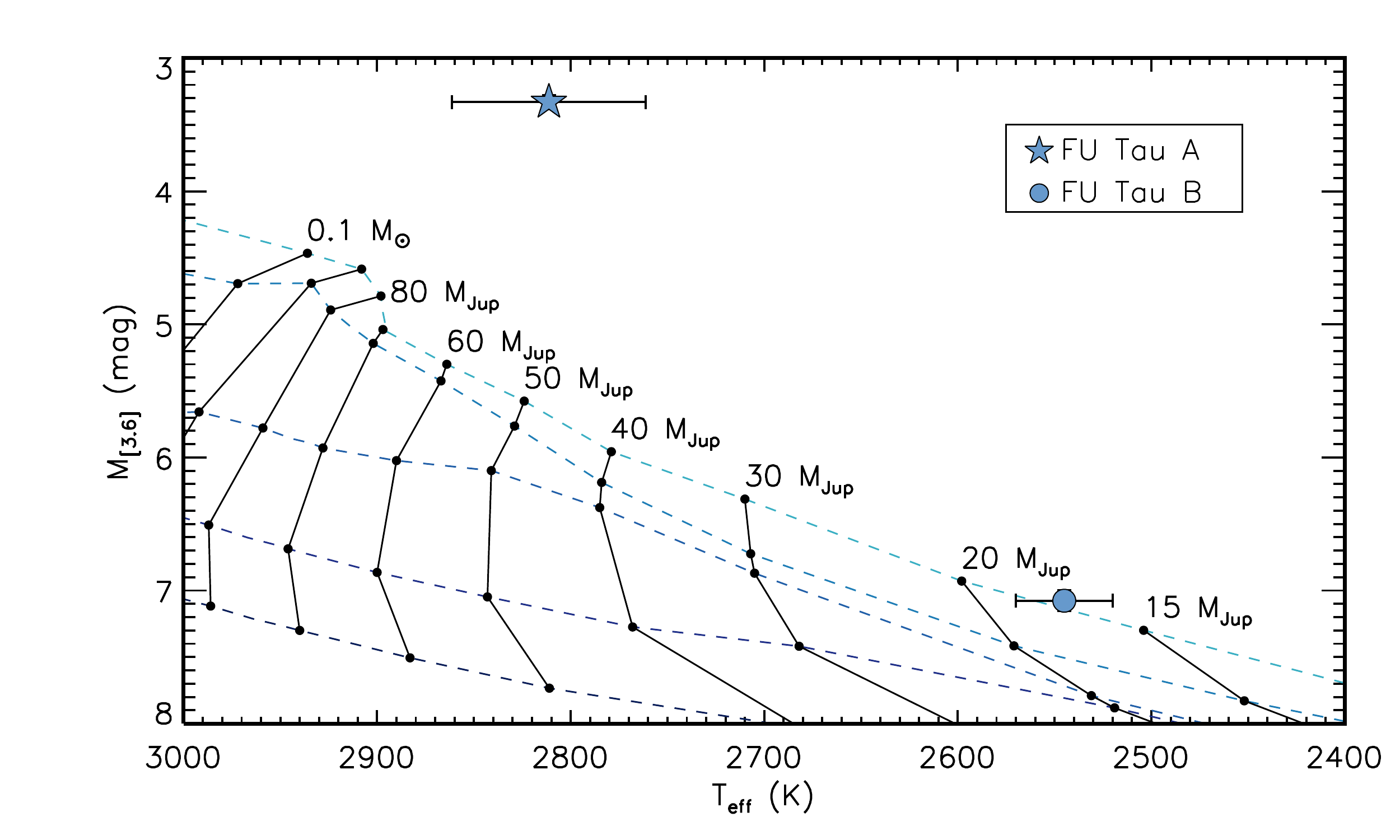}
\caption{H-R diagram for FU Tau AB. For both components, $T_{\mathrm{eff}}$ was determined using the temperature scale for young stars from \citet{herczeg14}. The 1, 2, 5, 10, and 20 Myr isochrones of \citet{allard12} are plotted (dashed lines) with mass tracks (solid lines) from 15 $M_{\mathrm{Jup}}$ to 0.1 $M_\odot$. FU Tau A is significantly overluminous compared to the 1 Myr isochrone.}
\label{hrd_futau}
\end{figure}

More recent observations of the primary suggest its spectral type is slightly earlier at $\sim$M6.5$-$M7 (e.g., \citealt{scholz12,herczeg14}). To assess whether the overluminosity of the primary continues into the IRAC channels, we converted the spectral types of FU Tau A (M6.5) and B (M9.25) to effective temperatures using the relation from \citet{herczeg14}. We also adopt a distance of 131.2 pc from \citet{gaia18} which is closer than the 140 pc distance that had been assumed in previous studies. The significant overluminosity of the primary persists across all IRAC channels and we show the example for Channel 1 in Figure \ref{hrd_futau}. FU Tau A is above the BT-Settl 1 Myr isochrone of \citet{allard12} by 2.36, 2.33, 2.31, and 2.24 mag in the IRAC channels. These results are of similar magnitude to the $\sim2$ mag overluminosity found in the $J$ band by \citet{scholz12}, though we do not confirm their claim of overluminosity for FU Tau B.

\subsubsection{FW Tau}
\label{fwtau1}
FW Tau AB is a close binary system of young M5.5 stars that harbors a third component (hereafter FW Tau C) at $\rho=2\farcs3$. The nature of FW Tau C is a matter of debate since different aspects of its observations are consistent with either a substellar companion surrounded by an edge-on disk or a PMC embedded in a low inclination disk (e.g., \citealt{kraus15,caceres15,wu17a}). \citet{kraus14} first confirmed FW Tau C as a comoving wide-separation companion with near-infrared observations using the Keck-II 10m telescope and NIRC2, and given its luminosity they estimated its mass to be $10\pm4$\,$M_{\mathrm{Jup}}$ for system ages between 1-5 Myr \citep{chabrier00b}. More recent ALMA Cycle 3 observations and modeling performed by \citet{wu17a} place the dynamical mass of FW Tau C closer to $\sim 0.1$\,$M_{\odot}$, but with a moderate disk inclination that does not explain its faint luminosity.

We are able to resolve FW Tau C and measure its mid-infrared photometry in all four IRAC channels. We measure the mid-infrared excess of FW Tau C to be $[3.6]-[8.0]=0.91\pm0.10$ mag. The 8 $\mu$m flux offers the prospect of an independent test between the proposed explanations, as the predicted SED for each case varies most at 8--12 $\mu$m. We discuss this comparison further in Section \ref{discussion}.

\subsubsection{USco 1610--2502}
2MASS J16101918--2502301 (hereafter USco 1610--2502) is a member of Upper Sco \citep{preibisch98} with a spectral type of M1. Its companion, USco 1610--2502 B, was confirmed to be a comoving companion by \citet{aller13}, who measured a spectral type of M5.5 and a projected separation of $\rho=5\farcs1$.

\begin{figure}
\centering
\includegraphics[trim={0 0 0 0},width=0.35\textwidth,angle=90]{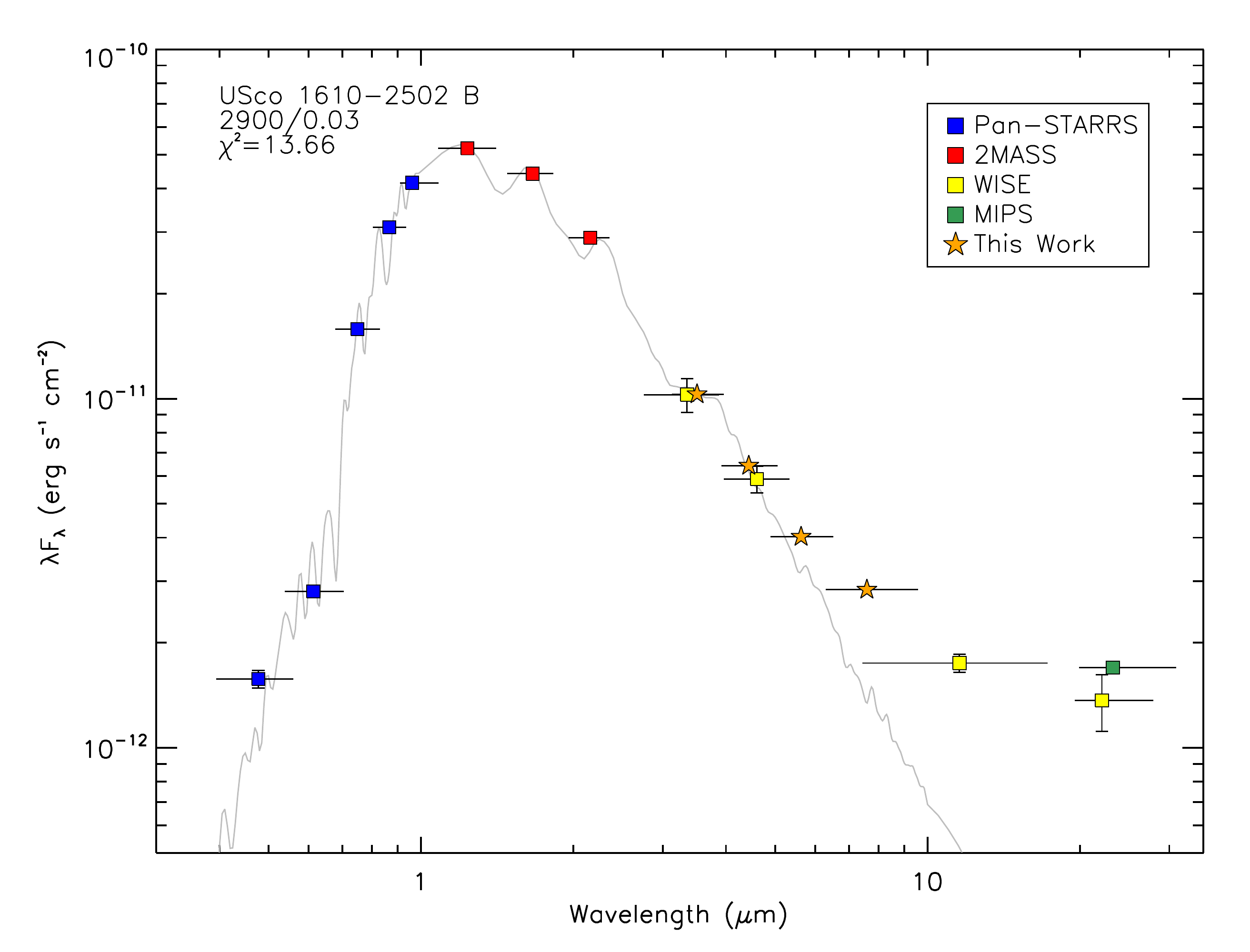}
\caption{Spectral energy distribution of USco 1610--2502 B. Photometric uncertainties are smaller than the symbol sizes if not shown. An $E(B-V)=0.03\pm0.025$ mag, $T_\mathrm{eff}=2900\pm50$ K BT-Settl model (gray) best fits the photometry of the primary for wavelengths between 0.4 and 3.6 $\mu$m. The bandpasses of the photometric filters are plotted as horizontal lines. The excess flux seen for wavelengths $>4.5$ $\mu$m is indicative of an irradiated disk with some grain growth and grain settling toward the midplane.}
\label{sed_usco1610}
\end{figure}

We have analyzed the SED of USco 1610--2502 B in a similar fashion as for the [SCH06] J0359+2009 system. We fit all PS1 and 2MASS photometry from the literature and include \textit{Spitzer}/IRAC $[3.6]$ and \textit{WISE} Band 1 photometry. The best-fitting model for USco 1610--2502 B is $T_{\mathrm{eff}}=2900\pm50$ K and $E(B-V)=0.03\pm0.025$ mag. Figure \ref{sed_usco1610} shows the best fitting model for USco 1610--2502 B as well as available photometry between 0.4 and 24 $\mu$m. There is no evidence for a circumstellar disk surrounding the primary but USco 1610--2502 B has a significant excess of $[3.6]-[8.0]=1.08\pm0.03$ mag. USco 1610--2502B also has 2MASS $K_S - [3.6]=0.36\pm0.06$ mag which is consistent with the colors of young stellar photospheres \citep{luhman10}. This confirms the ``transitional disk" designation from \citet{luhman12}. \citet{barenfeld16} measured a 0.88 mm continuum flux of $0.30\pm0.14$ mJy and estimate an upper limit on the dust disk mass to be $<$0.5 $M_{\oplus}$. 

\citet{cieza07} introduced a two-parameter scheme to understand the SED morphology of transitional disks based on identifying the longest wavelength at which the observed flux is dominated by stellar photosphere, $\lambda_{turnoff}$, and the slope of the IR excess, $\alpha_{excess}$, between $\lambda_{turnoff}$ and 24 $\mu$m. In this classification system, $\lambda_{turnoff}$ corresponds to the dust temperature, and therefore size, of the inner hole, while $\alpha_{excess}$ indicates the sharpness of the inner hole. Disks completely cleared of inner-hole dust have large and positive $\alpha_{excess}$ values, while disks undergoing significant grain growth and dust settling have large, negative $\alpha_{excess}$ values \citep{dullemond04}. For USco 1610--2502 B, $\lambda_{turnoff}=5.8$ $\mu$m and $\alpha_{excess}=-0.61$, indicative of an irradiated disk with some grain growth and grain settling toward the midplane. No mid-infrared spectroscopy exists for USco 1610--2502 B so it is not yet possible to constrain its inner hole size in the absence of millimeter imaging \citep{espaillat12}.

\subsection{Detection Limits}
\label{dls}
\begin{figure}
\centering
\includegraphics[trim={2.5cm 3.5cm 3.5cm 4.5cm},width=0.4\textwidth]{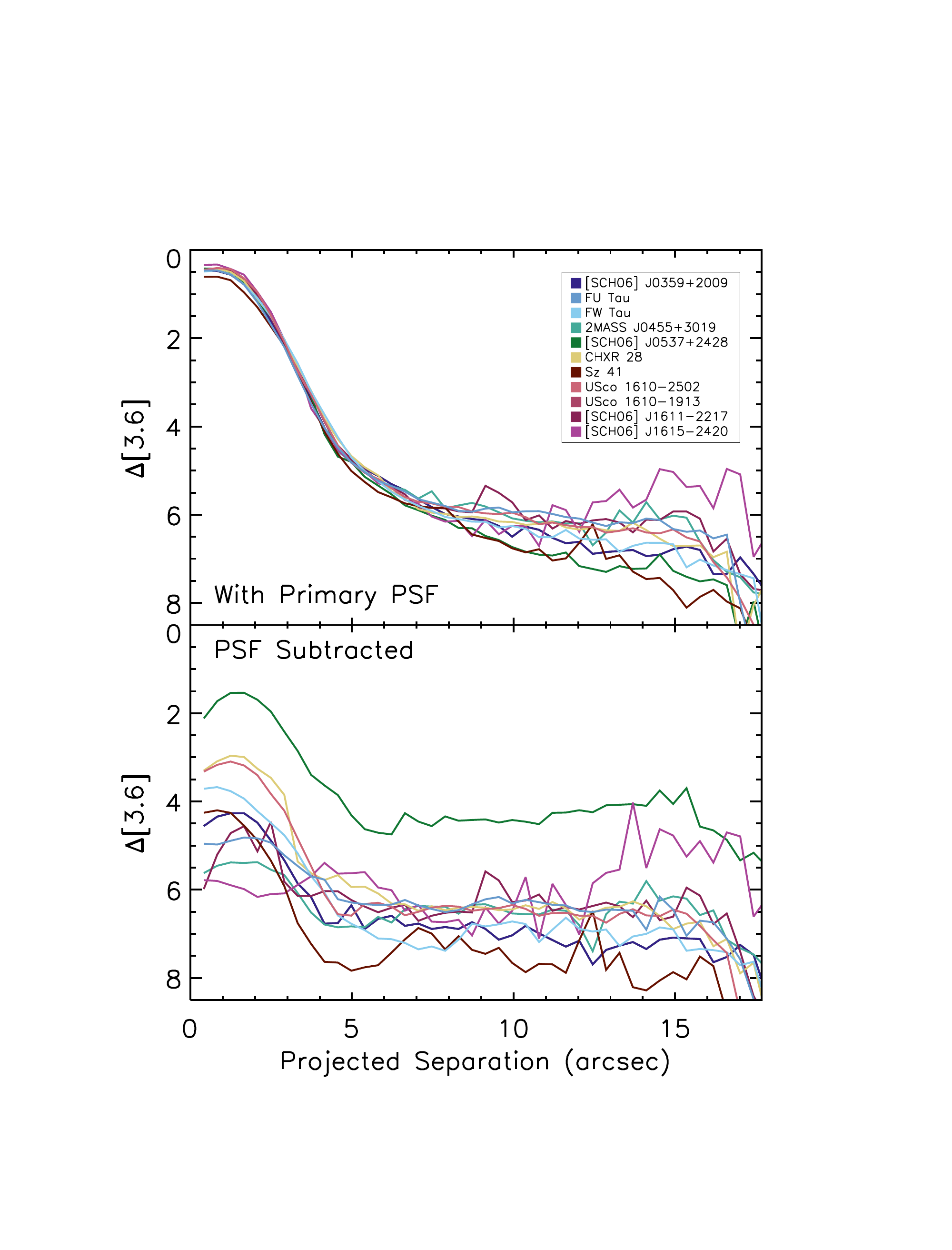}
\caption{Contrast limits for the stacked IRAC Channel 1 images of our target sample. The top panel shows the contrast curves prior to PSF subtraction as a function of projected separation from the primary star in arcseconds. The bottom panel shows the corresponding contrast curve once the primary PSF has been subtracted.}
\label{dl_i1}
\end{figure}

\begin{figure}
\centering
\includegraphics[trim={2.5cm 3.5cm 3.5cm 4.5cm},width=0.4\textwidth]{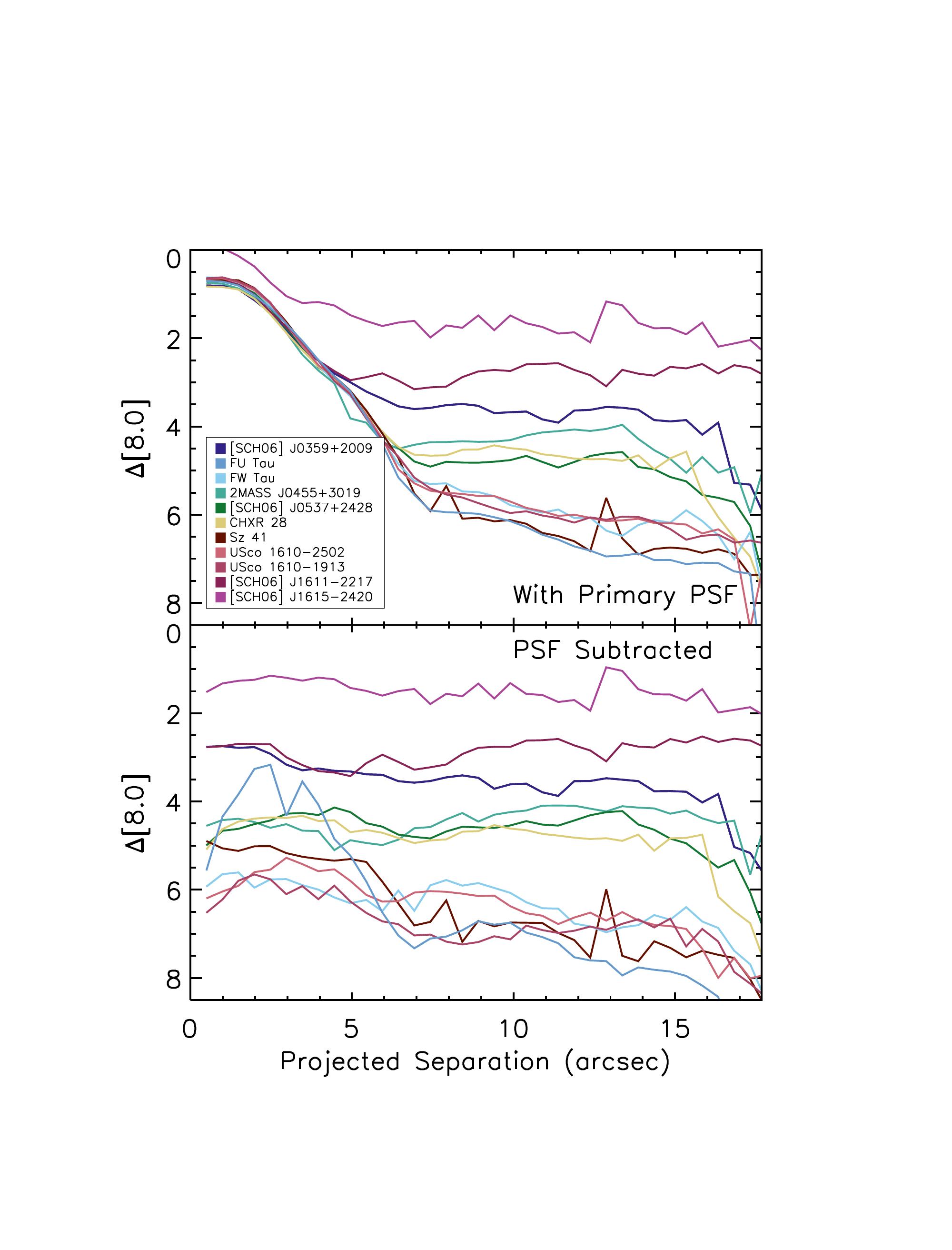}
\caption{Same as Figure \ref{dl_i1} but for IRAC Channel 4.}
\label{dl_i4}
\end{figure}

We evaluated the sensitivity to low-mass companions in terms of contrast for each target system individually by image and channel as a function of radial separation from the primary star. We performed aperture photometry by measuring the flux inside 100 randomly drawn apertures of radius 1 FWHM at each radius from the central star. We calculated the mean and standard deviation of these 100 fluxes to find the limiting flux and converted this value into \textit{Spitzer}/IRAC magnitudes to obtain 3-$\sigma$ limits. We also evaluated the sensitivity to low-mass companions in terms of contrast for each target system with the central star subtracted from the image in the same manner. The results of these calculations are presented as contrast curves in Figure \ref{dl_i1} and Figure \ref{dl_i4} for Channels 1 and 4, respectively. The curves show that the detectable contrast between central star and companion grows with increasing distance from the primary. For the images without primary PSF subtraction, a plateau is reached after $\sim 8\farcs5$ which corresponds to physical separations of 1000 au for the closest member of our target sample (114.5 pc) and 1800 au for the furthest (213.1 pc). Detectable contrast limits inward of $8\farcs5$ improve after PSF subtraction by an average of 5.5 magnitudes in Channel 1 and to the background noise-limit in Channel 4 for the majority of our target sample.

We convert our contrast curves into limiting masses using the BT-Settl evolutionary models of \citet{allard12} at 5 Myr and the measured absolute magnitudes of our target sample in each IRAC Channel. The contrast and mass limits reached as a function of radial separation from the PSF-subtracted images are presented in Tables \ref{cl_tab} and \ref{ml_tab}, respectively. The 5 Myr BT-Settl evolutionary model does not go below 10 $M_{\mathrm{Jup}}$ ($M_{[3.6]}>9.227$ mag) so we do not quote lower mass companions even though we are sensitive to them. Using the older COND-based of models of \citet{allard01} we find we are sensitive to companion masses as low as 0.8 $M_{\mathrm{Jup}}$ at $10\arcsec$ away from a 1 Myr system and 2.9 $M_{\mathrm{Jup}}$ at $10\arcsec$ away from a 10 Myr system.

\section{Discussion}
\label{discussion}

\subsection{Optimizing the Search for Wide PMCs with Spitzer}
The under-sampled PSF of \textit{Spitzer}/IRAC has made detection of exoplanets in the mid-infrared difficult, especially at small angular separations ($\sim\lambda /D$) that most closely approach solar-system scales. Here, we have shown that our framework to model the IRAC PSF is successful at recovering known or candidate companions at a wide range of projected separations from their hosts. Previous analyses of archival \textit{Spitzer}/IRAC images have placed initial constraints on the frequency of gas giant companions on wide orbits. \citet{durkan16} find companions with 0.5--13 $M_{\mathrm{Jup}}$ at separations of 100--1000 au occur with an upper frequency limit of 9\% based on a sample of 121 stars. \citet{baron18} probe further separations of 1000--5000 au in their 177 star sample, finding an occurrence rate of $<3$\% for 1--13 $M_{\mathrm{Jup}}$ companions. As mentioned previously, both surveys searched for wide companion systems in young moving groups that are closer ($<100$ pc) and older ($>10$ Myr) than the regions from which our target sample was created. Because of this, the contrast between primary star and companion is more severe in those surveys. We have shown that our pipeline can measure photometry at a few $\lambda /D$ in 1--10 Myr star-forming regions that are $>100$ pc away. Our pipeline will enable a systematic exploration of the demographics and properties (e.g., companion mass functions, semi-major axis distributions, disk frequencies) of wide-orbit, low-mass companions systems for samples of discrete stars in future work.

Nine systems in our target sample have had their IRAC photometry measured previously. When comparing our measurements to the latest IRAC measurement reported almost all agree within the errors. Exceptions to this are the brightest systems in our sample whose multiplicity was determined with AO imaging after the IRAC photometric measurements were reported (e.g., CHXR 28, Sz 41) or primaries with significant variability between epochs (FU Tau A). In addition, Sz 41 A is saturated in Channel 4 as seen in Figure \ref{stacked_residuals} by the characteristic ringed appearance in the PSF subtraction residuals. We are still able to resolve the companion and measure its photometry. The measured total flux from our pipeline agrees with the previous reported flux.

The presence of bright, unassociated stars in or just outside the $25\times 25$ fitting image mostly does not hinder the quality of the measured photometry significantly for our sample. One exception, 2MASS J04554970+3019400, is $\sim 30\arcsec$ away from HD 31305, a $V=7.6$ mag A0 star. While its primary photometry is unaffected, the photometry of the neighbor is contaminated by stray light that is not effectively modeled by the background parameter. In these cases, fitting and subtracting off the flux of the nearby bright star prior to PSF subtraction with our framework could allow the companion to be more accurately fit.

[SCH06] J0359+2009 and [SCH06] J0537+2428 were the only targets that had not had their IRAC photometry measured and reported in the literature. These targets were located at the outskirts of canonical star-forming regions and until recently not studied in detail. Now with \textit{Gaia} revealing so many more young systems, it is possible to evaluate whether they have wide companions and circumstellar disks, as well as re-evaluate interesting systems or systems where multiplicity was determined later.

More broadly, the Mid-Infrared Instrument (MIRI) on the \textit{James Webb Space Telescope} will provide key insight by constraining disk sizes for young planetary-mass companions known and discovered prior to its launch. Previous searches for disk emission at radio wavelengths have yielded mostly upper limits which suggests wide-orbit PMCs may have smaller, hotter disks and are more suited for characterization in the mid-infrared \citep{wu17b}. Our framework provides an efficient way to build targets of interest to be studied with \textit{JWST} and the ELTs coming on-line in the next decade.

\begin{figure}
\centering
\includegraphics[width=0.35\textwidth,angle=90]{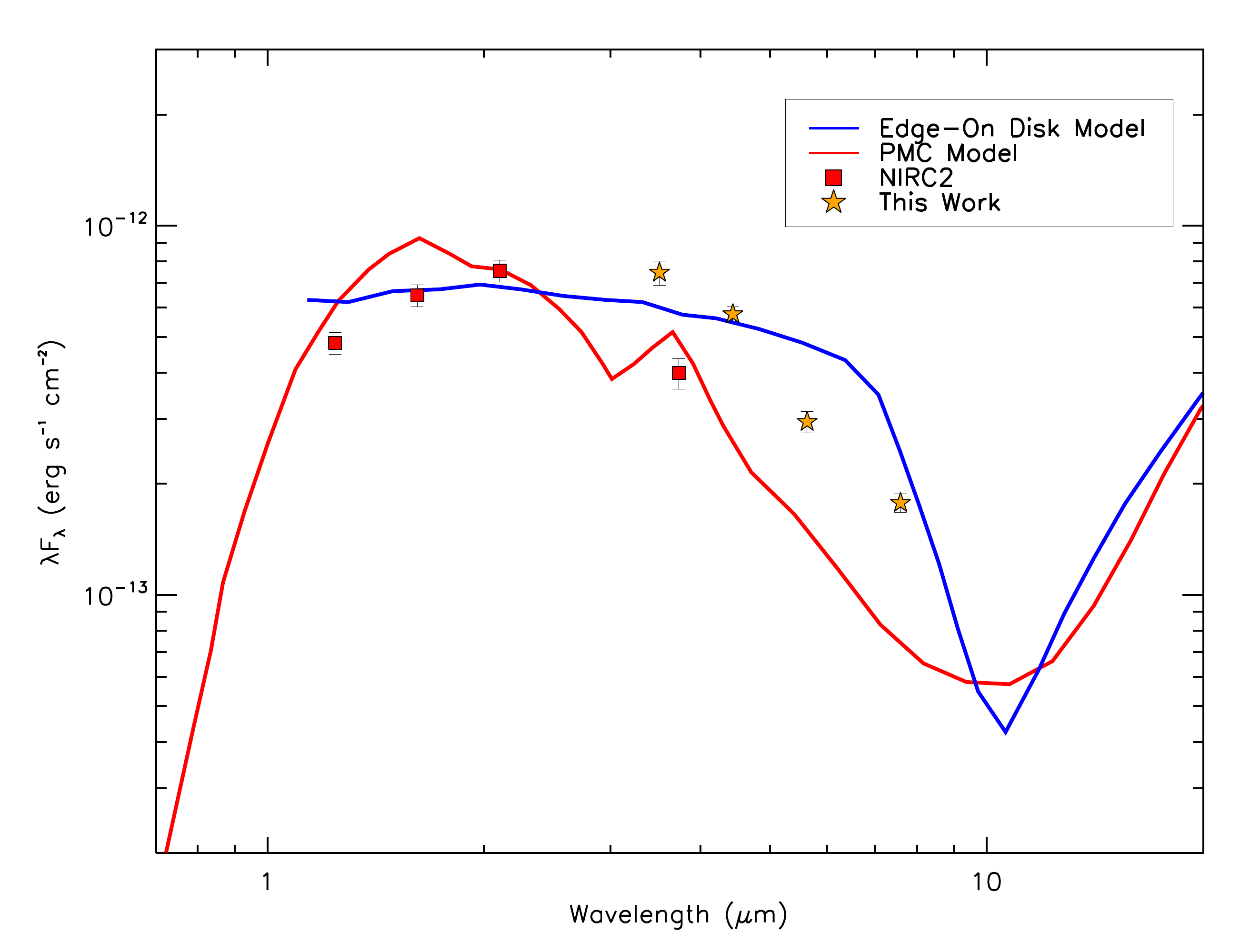}
\caption{SED for the two proposed scenarios for the nature of the FW Tau system. The blue and red lines are the brown-dwarf/edge-on-disk model and PMC model from \citet{caceres15}, respectively. The red squares are the NIRC2 photometric observations obtained by \citet{kraus14} while the orange stars are the \textit{Spitzer}/IRAC photometric measurements from this work. The brightness of the IRAC photometry is more consistent with the edge-on disk model in Channels 1, 2, and 4 but we note that the SED shape across the four IRAC channels appears to be more coincident with the PMC model. }
\label{fwtau_sed}
\end{figure}

\subsection{The Nature of FW Tau}
\label{fwtau2}
As we described in \ref{fwtau1}, two scenarios have been proposed to explain FW Tau C: a PMC embedded in a low inclination disk or a substellar companion surrounded by an edge-on disk. \citet{caceres15} constructed the FW Tau system SED from near-infrared to millimeter wavelengths to explore the possible interpretations via disk modeling. Although they did not come to a definitive conclusion about the nature of FW Tau C with the limited data available, their models differ most in SED shape and brightness in the mid-infrared from 3--10 $\mu$m (see Figure 3 of \citealt{caceres15}).

We can independently test the proposed scenarios because we have resolved FW Tau C across all IRAC channels. We find that the brightness of the IRAC photometry is more consistent with the edge-on disk model in Channels 1, 2, and 4 but also note that the SED shape across the four IRAC channels is actually more coincident with the PMC model (see Figure \ref{fwtau_sed}). Given the conflicting indications of this object's nature from across its SED, the enhanced sensitivity, spatial resolution, and wavelength coverage (3--30 $\mu$m) of MIRI  may ultimately resolve its true nature. 

\section{Summary}
We have developed an MCMC-based PSF fitter to re-analyze archival \textit{Spitzer}/IRAC images of 11 young, low-mass stars with varying spectral types that host faint confirmed or candidate companions at a range of projected separations. Our framework accurately models the flux of the system allowing us to measure the mid-infrared photometry for any astrophysical source in the vicinity of young stars of interest. We recover six confirmed, and two candidate low-mass companions, two of which have never had \textit{Spitzer}/IRAC photometry reported in the literature previously. One of these, [SCH06] J0359+2009 B, is a new companion with mass $20 \pm 5 M_{\mathrm{Jup}}$ and $[3.6]-[8.0]$ color indicative of a circum(sub)stellar disk. Using the evolutionary models of \citet{allard12}, we show that we are sensitive to companions $<10$\,$M_{\mathrm{Jup}}$ in the IRAC images. Our PSF-fitting framework is finally opening up a regime of parameter space that has yet to be studied in detail, revealing low-mass companions and whether they host disks.

\acknowledgments
We thank the referee for providing a helpful review that improved the clarity of our paper. R.A.M. acknowledges support from the Donald D.\ Harrington Doctoral Fellowship and the NASA Earth \& Space Science Fellowship. This work is based on observations made with the \textit{Spitzer Space Telescope}, which is operated by the Jet Propulsion Laboratory, California Institute of Technology under a contract with NASA. Support for this work was provided by an award issued by JPL/Caltech. This publication makes use of data products from the Two Micron All Sky Survey, which is a joint project of the University of Massachusetts and the Infrared Processing and Analysis Center/California Institute of Technology, funded by the National Aeronautics and Space Administration and the National Science Foundation. This publication makes use of data products from the Wide-field Infrared Survey Explorer, which is a joint project of the University of California, Los Angeles, and the Jet Propulsion Laboratory/California Institute of Technology, funded by the National Aeronautics and Space Administration. The Pan-STARRS1 Surveys (PS1) and the PS1 public science archive have been made possible through contributions by the Institute for Astronomy, the University of Hawaii, the Pan-STARRS Project Office, the Max-Planck Society and its participating institutes, the Max Planck Institute for Astronomy, Heidelberg and the Max Planck Institute for Extraterrestrial Physics, Garching, The Johns Hopkins University, Durham University, the University of Edinburgh, the Queen's University Belfast, the Harvard-Smithsonian Center for Astrophysics, the Las Cumbres Observatory Global Telescope Network Incorporated, the National Central University of Taiwan, the Space Telescope Science Institute, the National Aeronautics and Space Administration under Grant No. NNX08AR22G issued through the Planetary Science Division of the NASA Science Mission Directorate, the National Science Foundation Grant No. AST-1238877, the University of Maryland, Eotvos Lorand University (ELTE), the Los Alamos National Laboratory, and the Gordon and Betty Moore Foundation.

\begin{deluxetable*}{lcccccccccccccc}
\tablecaption{Companion Contrast Limits}
\tablehead{Target & Distance & Ch. & $M$ &Exp.\ Time & \multicolumn{10}{c}{Contrast (mag) at $\rho=$(arcsec)} \\ \cline{6-15} \\ \multicolumn{15}{c}{\vspace{-0.5cm}} \\
 & (pc) & & (mag) & (s) & 0.5 & 1.0 & 1.5 & 2.0 & 3.0 & 4.0 & 5.0 & 7.0 & 10.0 & 12.0}
\startdata
2MASS J03590986+2009361	& 117.4	& 1	& 6.85	& 10.4	& 4.51	& 4.31	& 4.27	& 4.44	& 5.45	& 6.55	& 6.38	& 6.77	& 7.01	& 7.17	\\
						&		& 2	& 6.73	& 10.4	& 4.98	& 4.80	& 4.74	& 4.92	& 5.48	& 5.83	& 5.89	& 5.99	& 6.33	& 6.03	\\
						&		& 3	& 6.70	& 10.4	& 3.52	& 3.57	& 3.63	& 3.77	& 4.23	& 4.42	& 4.53	& 4.29	& 4.26	& 4.34	\\
						&		& 4	& 6.65	& 10.4	& 2.76	& 2.75	& 2.78	& 2.77	& 3.17	& 3.26	& 3.33	& 3.57	& 3.61	& 3.54	\\
2MASS J04233539+2503026	& 131.2	& 1	& 2.76	& 0.4		& 4.91	& 4.90	& 4.78	& 4.73	& 5.17	& 5.86	& 6.07	& 6.31	& 6.13	& 6.37	\\
						&		& 2	& 2.15	& 0.4		& 4.23	& 4.32	& 4.45	& 4.63	& 5.35	& 6.31	& 6.75	& 6.75	& 7.19	& 6.75	\\
						&		& 3	& 1.70	& 10.4	& 4.91	& 4.87	& 4.85	& 4.80	& 4.83	& 5.92	& 6.96	& 6.91	& 7.23	& 7.44	\\
						&		& 4	& 0.94	& 10.4	& ...		& 5.61	& 4.22	& 3.54	& 4.30	& 4.27	& 5.40	& 7.14	& 6.89	& 7.54	\\
2MASS J04292971+2616532	& 145	& 1	& 3.20	& 1.0		& 6.87	& 6.46	& 6.10	& 5.95	& 5.93	& 6.75	& 8.07	& 8.27	& 7.65	& 7.66	\\
						&		& 2	& 3.09	& 1.0		& 6.22	& 5.71	& 5.44	& 5.35	& 5.65	& 6.81	& 6.77	& 6.83	& 6.67	& 6.69	\\
						&		& 3	& 3.03	& 10.4	& 6.52	& 6.37	& 6.40	& 6.53	& 7.18	& 7.68	& 8.38	& 6.98	& 6.66	& 6.74	\\
						&		& 4	& 3.04	& 10.4	& 5.75	& 5.62	& 5.56	& 5.78	& 5.72	& 5.97	& 6.19	& 5.75	& 5.92	& 6.80	\\
2MASS J04554970+3019400	& 156	& 1	& 5.37	& 10.4	& 5.81	& 5.64	& 5.35	& 5.20	& 5.60	& 6.45	& 6.81	& 6.44	& 6.40	& 6.11	\\
						&		& 2	& 5.18	& 10.4	& 6.28	& 5.81	& 5.51	& 5.31	& 5.65	& 6.39	& 6.67	& 6.23	& 6.57	& 5.93	\\
						&		& 3	& 4.96	& 10.4	& 4.74	& 4.37	& 4.30	& 4.43	& 4.90	& 5.12	& 5.00	& 4.83	& 4.48	& 4.11	\\
						&		& 4	& 4.61	& 10.4	& 4.40	& 4.50	& 4.63	& 4.61	& 4.35	& 4.14	& 4.20	& 4.58	& 3.99	& 3.68	\\
2MASS J05373850+2428517	& 114.5	& 1	& 5.12	& 10.4	& 2.04	& 1.65	& 1.54	& 1.67	& 2.52	& 3.55	& 4.33	& 4.43	& 4.42	& 4.20	\\
						&		& 2	& 5.04	& 10.4	& 6.31	& 5.96	& 5.55	& 5.34	& 5.45	& 6.24	& 7.16	& 7.02	& 6.69	& 6.70	\\
						&		& 3	& 4.96	& 10.4	& 4.60	& 4.50	& 4.58	& 4.83	& 5.65	& 5.77	& 5.65	& 5.66	& 5.53	& 5.53	\\
						&		& 4	& 4.94	& 10.4	& 4.99	& 4.66	& 4.62	& 4.51	& 4.28	& 4.29	& 4.27	& 4.81	& 4.52	& 4.40	\\
2MASS J11075588--7727257	& 202.1	& 1	& 1.46	& 0.4		& 3.21	& 3.07	& 3.00	& 3.09	& 4.13	& 5.03	& 5.45	& 6.46	& 6.44	& 6.53	\\
						&		& 2	& 1.47	& 0.4		& 3.92	& 3.82	& 3.69	& 3.60	& 3.48	& 4.49	& 4.77	& 5.63	& 5.85	& 5.74	\\
						&		& 3	& 1.27	& 0.4		& 4.66	& 4.35	& 4.15	& 4.14	& 4.32	& 4.49	& 4.85	& 5.20	& 4.63	& 4.48	\\
						&		& 4	& 1.30	& 0.4		& 5.11	& 4.64	& 4.48	& 4.38	& 4.36	& 4.40	& 4.50	& 4.82	& 4.57	& 4.81	\\
2MASS J11122441--7637064	& 192.7	& 1	& 0.89	& 0.4		& 4.57	& 4.49	& 4.65	& 4.89	& 6.00	& 6.97	& 7.81	& 6.73	& 7.67	& 6.79	\\
						&		& 2	& 0.46	& 0.4		& 7.32	& 5.83	& 5.47	& 5.63	& 6.91	& 7.69	& 7.66	& 7.10	& 8.12	& 7.99	\\
						&		& 3	& -0.03	& 0.4		& 4.94	& 4.78	& 4.77	& 4.84	& 5.24	& 5.66	& 6.08	& 6.10	& 5.58	& 5.94	\\
						&		& 4	& -1.07	& 0.4		& 4.88	& 4.95	& 5.12	& 4.97	& 5.18	& 5.25	& 5.38	& 6.78	& 6.73	& 7.34	\\
2MASS J16101918--2502301	& 152	& 1	& 2.36	& 0.4		& 4.95	& 4.84	& 4.80	& 4.79	& 5.41	& 6.33	& 7.11	& 6.58	& 6.59	& 6.66	\\
						&		& 2	& 2.42	& 0.4		& 5.03	& 4.74	& 4.65	& 4.71	& 5.11	& 5.38	& 5.67	& 5.23	& 5.77	& 5.49	\\
						&		& 3	& 2.34	& 10.4	& 4.60	& 4.61	& 4.54	& 4.54	& 4.95	& 5.41	& 5.79	& 6.44	& 6.45	& 6.46	\\
						&		& 4	& 2.28	& 10.4	& 6.63	& 6.50	& 6.28	& 5.90	& 5.49	& 5.50	& 6.02	& 6.33	& 6.32	& 6.76	\\
2MASS J16103196--1913062	& 132.9	& 2	& 3.06	& 1.2		& 9.59	& 7.36	& 6.66	& 6.43	& 6.63	& 7.23	& 7.53	& 7.31	& 7.27	& 7.34	\\
						&		& 4	& 2.96	& 10.4	& 5.75	& 5.49	& 5.36	& 5.16	& 5.15	& 5.67	& 6.18	& 7.00	& 7.63	& 7.28	\\
2MASS J16111711--2217173	& 213.1	& 1	& 6.22	& 10.4	& 5.82	& 5.00	& 4.62	& 5.03	& 5.89	& 6.07	& 6.24	& 6.66	& 6.28	& 6.44	\\
						&		& 2	& 6.11	& 10.4	& 5.37	& 4.99	& 4.72	& 6.16	& 6.08	& 5.51	& 5.72	& 5.83	& 6.08	& 5.72	\\
						&		& 3	& 5.70	& 10.4	& 4.78	& 4.47	& 4.38	& 4.22	& 3.96	& 3.81	& 3.71	& 3.76	& 3.54	& 3.49	\\
						&		& 4	& 6.08	& 10.4	& 2.77	& 2.75	& 2.69	& 2.70	& 3.01	& 3.31	& 3.40	& 3.27	& 2.73	& 2.76	\\
2MASS J16151116--2420153	& 143.8	& 1	& 7.00	& 10.4	& 5.79	& 5.84	& 5.95	& 6.13	& 6.04	& 5.50	& 5.63	& 6.37	& 6.41	& 6.96	\\
						&		& 2	& 6.92	& 10.4	& 4.84	& 4.88	& 5.32	& 6.16	& 7.69	& 8.43	& 6.03	& 6.25	& 5.61	& 4.96	\\
						&		& 3	& 6.85	& 10.4	& 3.53	& 3.37	& 3.36	& 3.44	& 3.96	& 4.23	& 4.17	& 4.14	& 3.41	& 3.30	\\
						&		& 4	& 6.90	& 10.4	& 1.52	& 1.32	& 1.26	& 1.24	& 1.20	& 1.20	& 1.43	& 1.50	& 1.37	& 1.76
\enddata	
\label{cl_tab}
\end{deluxetable*}

\begin{deluxetable*}{lcccccccccccccc}
\tablecaption{Companion Mass Limits}
\tablehead{Target & Distance & Ch. & $M$ &Exp.\ Time & \multicolumn{10}{c}{Mass Limit ($M_{\mathrm{Jup}}$) at $\rho=$(arcsec)} \\ \cline{6-15} \\ \multicolumn{15}{c}{\vspace{-0.5cm}} \\
 & (pc) & & (mag) & (s) & 0.5 & 1.0 & 1.5 & 2.0 & 3.0 & 4.0 & 5.0 & 7.0 & 10.0 & 12.0}
\startdata
2MASS J03590986+200936	& 117.4	& 1	& 6.84	& 10.4	& $<10$	& $<10$	& $<10$	& $<10$	& $<10$	& $<10$	& $<10$	& $<10$	& $<10$	& $<10$	\\
						&		& 2	& 6.73	& 10.4	& $<10$	& $<10$	& $<10$	& $<10$	& $<10$	& $<10$	& $<10$	& $<10$	& $<10$	& $<10$	\\
						&		& 3	& 6.76	& 10.4	& $<10$	& $<10$	& $<10$	& $<10$	& $<10$	& $<10$	& $<10$	& $<10$	& $<10$	& $<10$	\\
						&		& 4	& 6.79	& 10.4	& $<10$	& $<10$	& $<10$	& $<10$	& $<10$	& $<10$	& $<10$	& $<10$	& $<10$	& $<10$	\\
2MASS J04233539+2503026	& 131.2	& 1	& 2.76	& 0.4		& 21		& 21		& 23		& 23		& 19		& 14		& 13		& 11		& 12		& 11		\\
						&		& 2	& 2.15	& 0.4		& 37		& 35		& 33		& 29		& 22		& 15		& 12		& 12		& $<10$	& 12		\\
						&		& 3	& 1.70	& 10.4	& 31		& 32		& 32		& 33		& 33		& 20		& 13		& 13		& 11		& $<10$	\\
						&		& 4	& 0.94	& 10.4	& ...		& 31		& 140	& 240	& 130	& 130	& 35		& 16		& 18		& 13		\\
2MASS J04292971+2616532	& 145	& 1	& 3.20	& 1.0		& $<10$	& $<10$	& $<10$	& 11		& 11		& $<10$	& $<10$	& $<10$	& $<10$	& $<10$	\\
						&		& 2	& 3.09	& 1.0		& $<10$	& 13		& 14		& 15		& 13		& $<10$	& $<10$	& $<10$	& $<10$	& $<10$	\\
						&		& 3	& 3.03	& 10.4	& $<10$	& $<10$	& $<10$	& $<10$	& $<10$	& $<10$	& $<10$	& $<10$	& $<10$	& $<10$	\\
						&		& 4	& 3.04	& 10.4	& 10		& 12		& 12		& 10		& 11		& $<10$	& $<10$	& 10		& $<10$	& $<10$	\\
2MASS J04554970+3019400	& 156	& 1	& 5.37	& 10.4	& $<10$	& $<10$	& $<10$	& $<10$	& $<10$	& $<10$	& $<10$	& $<10$	& $<10$	& $<10$	\\
						&		& 2	& 5.18	& 10.4	& $<10$	& $<10$	& $<10$	& $<10$	& $<10$	& $<10$	& $<10$	& $<10$	& $<10$	& $<10$	\\
						&		& 3	& 4.96	& 10.4	& $<10$	& $<10$	& $<10$	& $<10$	& $<10$	& $<10$	& $<10$	& $<10$	& $<10$	& $<10$	\\
						&		& 4	& 4.61	& 10.4	& $<10$	& $<10$	& $<10$	& $<10$	& $<10$	& 11		& 10		& $<10$	& 12		& 14		\\
2MASS J05373850+2428517	& 114.5	& 1	& 5.12	& 10.4	& 27		& 32		& 34		& 32		& 22		& 14		& $<10$	& $<10$	& $<10$	& $<10$	\\
						&		& 2	& 5.04	& 10.4	& $<10$	& $<10$	& $<10$	& $<10$	& $<10$	& $<10$	& $<10$	& $<10$	& $<10$	& $<10$	\\
						&		& 3	& 4.96	& 10.4	& $<10$	& $<10$	& $<10$	& $<10$	& $<10$	& $<10$	& $<10$	& $<10$	& $<10$	& $<10$	\\
						&		& 4	& 4.94	& 10.4	& $<10$	& $<10$	& $<10$	& $<10$	& $<10$	& $<10$	& $<10$	& $<10$	& $<10$	& $<10$	\\
2MASS J11075588--7727257	& 202.1	& 1	& 1.46	& 0.4		& 240 	& 260	& 280	& 260	& 110	& 38		& 30		& 19		& 19		& 19		\\
						&		& 2	& 1.47	& 0.4		& 120	& 130	& 150	& 160	& 170	& 50		& 40		& 26		& 24		& 25		\\
						&		& 3	& 1.27	& 0.4		& 49		& 78		& 110	& 110	& 81		& 71		& 42		& 34		& 50		& 69		\\
						&		& 4	& 1.30	& 0.4		& 34		& 46		& 58		& 70		& 71		& 67		& 56		& 40		& 49		& 40		\\
2MASS J11122441--7637064	& 192.7	& 1	& 0.89	& 0.4		& 130	& 140	& 120	& 80		& 30		& 20		& 14		& 22		& 15		& 21		\\
						&		& 2	& 0.46	& 0.4		& 19		& 39		& 57		& 46		& 23		& 17		& 17		& 21		& 14		& 15		\\
						&		& 3	& -0.03	& 0.4		& 170	& 190	& 200	& 190	& 140	& 78		& 45		& 44		& 84		& 50		\\
						&		& 4	& -1.07	& 0.4		& 420	& 390	& 340	& 390	& 330	& 300	& 280	& 67		& 72		& 37		\\
2MASS J16101918--2502301	& 152	& 1	& 2.36	& 0.4		& 25		& 26		& 27		& 27		& 20		& 14		& $<10$	& 12		& 12		& 11		\\
						&		& 2	& 2.42	& 0.4		& 22		& 25		& 26		& 26		& 21		& 19		& 17		& 20		& 16		& 18		\\
						&		& 3	& 2.34	& 10.4	& 27		& 27		& 28		& 28		& 23		& 19		& 16		& 12		& 12		& 12		\\
						&		& 4	& 2.28	& 10.4	& 10		& 12		& 12		& 15		& 18		& 18		& 14		& 12		& 12		& $<10$	\\
2MASS J16103196--1913062	& 132.9	& 2	& 3.06	& 1.2		& $<10$	& $<10$	& $<10$	& $<10$	& $<10$	& $<10$	& $<10$	& $<10$	& $<10$	& $<10$	\\
						&		& 4	& 2.96	& 10.4	& 11		& 13		& 14		& 16		& 16		& 12		& $<10$	& 12		& $<10$	& $<10$	\\
2MASS J16111711--2217173	& 213.1	& 1	& 6.22	& 10.4	& $<10$	& $<10$	& $<10$	& $<10$	& $<10$	& $<10$	& $<10$	& $<10$	& $<10$	& $<10$	\\
						&		& 2	& 6.11	& 10.4	& $<10$	& $<10$	& $<10$	& $<10$	& $<10$	& $<10$	& $<10$	& $<10$	& $<10$	& $<10$	\\
						&		& 3	& 5.70	& 10.4	& $<10$	& $<10$	& $<10$	& $<10$	& $<10$	& $<10$	& $<10$	& $<10$	& $<10$	& 10		\\
						&		& 4	& 6.08	& 10.4	& 10		& 10		& 11		& 11		& $<10$	& $<10$	& $<10$	& $<10$	& 10		& 10		\\
2MASS J16151116--2420153	& 143.8	& 1	& 7.00	& 10.4	& $<10$	& $<10$	& $<10$	& $<10$	& $<10$	& $<10$	& $<10$	& $<10$	& $<10$	& $<10$	\\
						&		& 2	& 6.92	& 10.4	& $<10$	& $<10$	& $<10$	& $<10$	& $<10$	& $<10$	& $<10$	& $<10$	& $<10$	& $<10$	\\
						&		& 3	& 6.85	& 10.4	& $<10$	& $<10$	& $<10$	& $<10$	& $<10$	& $<10$	& $<10$	& $<10$	& $<10$	& $<10$	\\
						&		& 4	& 6.90	& 10.4	& 13		& 15		& 15		& 16		& 16		& 16		& 14		& 14		& 16		& 11
\enddata
\label{ml_tab}
\end{deluxetable*}

\clearpage
\bibliographystyle{aasjournal}
\bibliography{swc_i.bib}

\begin{thebibliography}{}
\expandafter\ifx\csname natexlab\endcsname\relax\def\natexlab#1{#1}\fi
\providecommand{\url}[1]{\href{#1}{#1}}

\bibitem[{{Allard} {et~al.}(2001){Allard}, {Hauschildt}, {Alexander},
  {Tamanai}, \& {Schweitzer}}]{allard01}
{Allard}, F., {Hauschildt}, P.~H., {Alexander}, D.~R., {Tamanai}, A., \&
  {Schweitzer}, A. 2001, \apj, 556, 357

\bibitem[{{Allard} {et~al.}(2012){Allard}, {Homeier}, \& {Freytag}}]{allard12}
{Allard}, F., {Homeier}, D., \& {Freytag}, B. 2012, Philosophical Transactions
  of the Royal Society of London Series A, 370, 2765

\bibitem[{{Aller} {et~al.}(2013){Aller}, {Kraus}, {Liu}, {Burgett}, {Chambers},
  {Hodapp}, {Kaiser}, {Magnier}, \& {Price}}]{aller13}
{Aller}, K.~M., {Kraus}, A.~L., {Liu}, M.~C., {et~al.} 2013, \apj, 773, 63

\bibitem[{{Andrews} {et~al.}(2013){Andrews}, {Rosenfeld}, {Kraus}, \&
  {Wilner}}]{andrews13}
{Andrews}, S.~M., {Rosenfeld}, K.~A., {Kraus}, A.~L., \& {Wilner}, D.~J. 2013,
  \apj, 771, 129

\bibitem[{{Bailer-Jones} {et~al.}(2018){Bailer-Jones}, {Rybizki}, {Fouesneau},
  {Mantelet}, \& {Andrae}}]{bailer-jones18}
{Bailer-Jones}, C.~A.~L., {Rybizki}, J., {Fouesneau}, M., {Mantelet}, G., \&
  {Andrae}, R. 2018, \aj, 156, 58

\bibitem[{{Bailey} {et~al.}(2014){Bailey}, {Meshkat}, {Reiter}, {Morzinski},
  {Males}, {Su}, {Hinz}, {Kenworthy}, {Stark}, {Mamajek}, {Briguglio}, {Close},
  {Follette}, {Puglisi}, {Rodigas}, {Weinberger}, \& {Xompero}}]{bailey14}
{Bailey}, V., {Meshkat}, T., {Reiter}, M., {et~al.} 2014, \apjl, 780, L4

\bibitem[{{Barenfeld} {et~al.}(2016){Barenfeld}, {Carpenter}, {Ricci}, \&
  {Isella}}]{barenfeld16}
{Barenfeld}, S.~A., {Carpenter}, J.~M., {Ricci}, L., \& {Isella}, A. 2016,
  \apj, 827, 142

\bibitem[{{Baron} {et~al.}(2018){Baron}, {Artigau}, {Rameau}, {Lafreni{\`e}re},
  {Gagn{\'e}}, {Malo}, {Albert}, {Naud}, {Doyon}, {Janson}, {Delorme}, \&
  {Beichman}}]{baron18}
{Baron}, F., {Artigau}, {\'E}., {Rameau}, J., {et~al.} 2018, \aj, 156, 137

\bibitem[{{Bate}(2005)}]{bate05}
{Bate}, M.~R. 2005, \mnras, 363, 363

\bibitem[{{Bodenheimer} \& {Burkert}(2001)}]{bodenheimer01}
{Bodenheimer}, P., \& {Burkert}, A. 2001, in IAU Symposium, Vol. 200, The
  Formation of Binary Stars, ed. H.~{Zinnecker} \& R.~{Mathieu}, 13

\bibitem[{{Boss}(1988)}]{boss88}
{Boss}, A.~P. 1988, \apj, 331, 370

\bibitem[{{Boss}(2011)}]{boss11}
---. 2011, \apj, 731, 74

\bibitem[{{Bowler}(2016)}]{bowler16}
{Bowler}, B.~P. 2016, \pasp, 128, 102001

\bibitem[{{Bowler} {et~al.}(2014){Bowler}, {Liu}, {Kraus}, \&
  {Mann}}]{bowler14}
{Bowler}, B.~P., {Liu}, M.~C., {Kraus}, A.~L., \& {Mann}, A.~W. 2014, \apj,
  784, 65

\bibitem[{{Bowler} {et~al.}(2011){Bowler}, {Liu}, {Kraus}, {Mann}, \&
  {Ireland}}]{bowler11}
{Bowler}, B.~P., {Liu}, M.~C., {Kraus}, A.~L., {Mann}, A.~W., \& {Ireland},
  M.~J. 2011, \apj, 743, 148

\bibitem[{{Boyd} \& {Whitworth}(2005)}]{boyd05}
{Boyd}, D.~F.~A., \& {Whitworth}, A.~P. 2005, \aap, 430, 1059

\bibitem[{{Bryan} {et~al.}(2016){Bryan}, {Bowler}, {Knutson}, {Kraus},
  {Hinkley}, {Mawet}, {Nielsen}, \& {Blunt}}]{bryan16}
{Bryan}, M.~L., {Bowler}, B.~P., {Knutson}, H.~A., {et~al.} 2016, \apj, 827,
  100

\bibitem[{{Caceres} {et~al.}(2015){Caceres}, {Hardy}, {Schreiber},
  {C{\'a}novas}, {Cieza}, {Williams}, {Hales}, {Pinte}, {M{\'e}nard}, \&
  {Wahhaj}}]{caceres15}
{Caceres}, C., {Hardy}, A., {Schreiber}, M.~R., {et~al.} 2015, \apjl, 806, L22

\bibitem[{{Chabrier} {et~al.}(2000){Chabrier}, {Baraffe}, {Allard}, \&
  {Hauschildt}}]{chabrier00b}
{Chabrier}, G., {Baraffe}, I., {Allard}, F., \& {Hauschildt}, P. 2000, \apj,
  542, 464

\bibitem[{{Chambers} {et~al.}(2016){Chambers}, {Magnier}, {Metcalfe},
  {Flewelling}, {Huber}, {Waters}, {Denneau}, {Draper}, {Farrow}, {Finkbeiner},
  {Holmberg}, {Koppenhoefer}, {Price}, {Saglia}, {Schlafly}, {Smartt},
  {Sweeney}, {Wainscoat}, {Burgett}, {Grav}, {Heasley}, {Hodapp}, {Jedicke},
  {Kaiser}, {Kudritzki}, {Luppino}, {Lupton}, {Monet}, {Morgan}, {Onaka},
  {Stubbs}, {Tonry}, {Banados}, {Bell}, {Bender}, {Bernard}, {Botticella},
  {Casertano}, {Chastel}, {Chen}, {Chen}, {Cole}, {Deacon}, {Frenk},
  {Fitzsimmons}, {Gezari}, {Goessl}, {Goggia}, {Goldman}, {Grebel}, {Hambly},
  {Hasinger}, {Heavens}, {Heckman}, {Henderson}, {Henning}, {Holman}, {Hopp},
  {Ip}, {Isani}, {Keyes}, {Koekemoer}, {Kotak}, {Long}, {Lucey}, {Liu},
  {Martin}, {McLean}, {Morganson}, {Murphy}, {Nieto-Santisteban}, {Norberg},
  {Peacock}, {Pier}, {Postman}, {Primak}, {Rae}, {Rest}, {Riess}, {Riffeser},
  {Rix}, {Roser}, {Schilbach}, {Schultz}, {Scolnic}, {Szalay}, {Seitz},
  {Shiao}, {Small}, {Smith}, {Soderblom}, {Taylor}, {Thakar}, {Thiel},
  {Thilker}, {Urata}, {Valenti}, {Walter}, {Watters}, {Werner}, {White},
  {Wood-Vasey}, \& {Wyse}}]{chambers16}
{Chambers}, K.~C., {Magnier}, E.~A., {Metcalfe}, N., {et~al.} 2016, ArXiv
  e-prints, arXiv:1612.05560

\bibitem[{{Cieza} {et~al.}(2007){Cieza}, {Padgett}, {Stapelfeldt}, {Augereau},
  {Harvey}, {Evans}, {Mer{\'{\i}}n}, {Koerner}, {Sargent}, {van Dishoeck},
  {Allen}, {Blake}, {Brooke}, {Chapman}, {Huard}, {Lai}, {Mundy}, {Myers},
  {Spiesman}, \& {Wahhaj}}]{cieza07}
{Cieza}, L., {Padgett}, D.~L., {Stapelfeldt}, K.~R., {et~al.} 2007, \apj, 667,
  308

\bibitem[{{Cutri} {et~al.}(2003){Cutri}, {Skrutskie}, {van Dyk}, {Beichman},
  {Carpenter}, {Chester}, {Cambresy}, {Evans}, {Fowler}, {Gizis}, {Howard},
  {Huchra}, {Jarrett}, {Kopan}, {Kirkpatrick}, {Light}, {Marsh}, {McCallon},
  {Schneider}, {Stiening}, {Sykes}, {Weinberg}, {Wheaton}, {Wheelock}, \&
  {Zacarias}}]{cutri03}
{Cutri}, R.~M., {Skrutskie}, M.~F., {van Dyk}, S., {et~al.} 2003, VizieR Online
  Data Catalog, 2246

\bibitem[{{de Zeeuw} {et~al.}(1999){de Zeeuw}, {Hoogerwerf}, {de Bruijne},
  {Brown}, \& {Blaauw}}]{deZeeuw99}
{de Zeeuw}, P.~T., {Hoogerwerf}, R., {de Bruijne}, J.~H.~J., {Brown}, A.~G.~A.,
  \& {Blaauw}, A. 1999, \aj, 117, 354

\bibitem[{{Dodson-Robinson} {et~al.}(2009){Dodson-Robinson}, {Veras}, {Ford},
  \& {Beichman}}]{dodson09}
{Dodson-Robinson}, S.~E., {Veras}, D., {Ford}, E.~B., \& {Beichman}, C.~A.
  2009, \apj, 707, 79

\bibitem[{{Dullemond} \& {Dominik}(2004)}]{dullemond04}
{Dullemond}, C.~P., \& {Dominik}, C. 2004, in Astronomical Society of the
  Pacific Conference Series, Vol. 321, Extrasolar Planets: Today and Tomorrow,
  ed. J.~{Beaulieu}, A.~{Lecavelier Des Etangs}, \& C.~{Terquem}, 361

\bibitem[{{Dupuy} {et~al.}(2010){Dupuy}, {Liu}, {Bowler}, {Cushing}, {Helling},
  {Witte}, \& {Hauschildt}}]{dupuy10}
{Dupuy}, T.~J., {Liu}, M.~C., {Bowler}, B.~P., {et~al.} 2010, \apj, 721, 1725

\bibitem[{{Durkan} {et~al.}(2016){Durkan}, {Janson}, \& {Carson}}]{durkan16}
{Durkan}, S., {Janson}, M., \& {Carson}, J.~C. 2016, \apj, 824, 58

\bibitem[{{Espaillat} {et~al.}(2012){Espaillat}, {Ingleby}, {Hern{\'a}ndez},
  {Furlan}, {D'Alessio}, {Calvet}, {Andrews}, {Muzerolle}, {Qi}, \&
  {Wilner}}]{espaillat12}
{Espaillat}, C., {Ingleby}, L., {Hern{\'a}ndez}, J., {et~al.} 2012, \apj, 747,
  103

\bibitem[{{Evans} {et~al.}(2009){Evans}, {Dunham}, {J{\o}rgensen}, {Enoch},
  {Mer{\'{\i}}n}, {van Dishoeck}, {Alcal{\'a}}, {Myers}, {Stapelfeldt},
  {Huard}, {Allen}, {Harvey}, {van Kempen}, {Blake}, {Koerner}, {Mundy},
  {Padgett}, \& {Sargent}}]{evans09}
{Evans}, II, N.~J., {Dunham}, M.~M., {J{\o}rgensen}, J.~K., {et~al.} 2009,
  \apjs, 181, 321

\bibitem[{{Fazio} {et~al.}(2004){Fazio}, {Hora}, {Allen}, {Ashby}, {Barmby},
  {Deutsch}, {Huang}, {Kleiner}, {Marengo}, {Megeath}, {Melnick}, {Pahre},
  {Patten}, {Polizotti}, {Smith}, {Taylor}, {Wang}, {Willner}, {Hoffmann},
  {Pipher}, {Forrest}, {McMurty}, {McCreight}, {McKelvey}, {McMurray}, {Koch},
  {Moseley}, {Arendt}, {Mentzell}, {Marx}, {Losch}, {Mayman}, {Eichhorn},
  {Krebs}, {Jhabvala}, {Gezari}, {Fixsen}, {Flores}, {Shakoorzadeh}, {Jungo},
  {Hakun}, {Workman}, {Karpati}, {Kichak}, {Whitley}, {Mann}, {Tollestrup},
  {Eisenhardt}, {Stern}, {Gorjian}, {Bhattacharya}, {Carey}, {Nelson},
  {Glaccum}, {Lacy}, {Lowrance}, {Laine}, {Reach}, {Stauffer}, {Surace},
  {Wilson}, {Wright}, {Hoffman}, {Domingo}, \& {Cohen}}]{fazio04}
{Fazio}, G.~G., {Hora}, J.~L., {Allen}, L.~E., {et~al.} 2004, \apjs, 154, 10

\bibitem[{{Feiden}(2016)}]{feiden16}
{Feiden}, G.~A. 2016, \aap, 593, A99

\bibitem[{{Fitzpatrick}(1999)}]{fitzpatrick99}
{Fitzpatrick}, E.~L. 1999, \pasp, 111, 63

\bibitem[{{Gaia Collaboration} {et~al.}(2018){Gaia Collaboration}, {Brown},
  {Vallenari}, {Prusti}, {de Bruijne}, {Babusiaux}, {Bailer-Jones}, {Biermann},
  {Evans}, {Eyer}, \& et~al.}]{gaia18}
{Gaia Collaboration}, {Brown}, A.~G.~A., {Vallenari}, A., {et~al.} 2018, \aap,
  616, A1

\bibitem[{{Ginski} {et~al.}(2014){Ginski}, {Schmidt}, {Mugrauer},
  {Neuh{\"a}user}, {Vogt}, {Errmann}, \& {Berndt}}]{ginski14}
{Ginski}, C., {Schmidt}, T.~O.~B., {Mugrauer}, M., {et~al.} 2014, \mnras, 444,
  2280

\bibitem[{{Gully-Santiago} {et~al.}(2017){Gully-Santiago}, {Herczeg},
  {Czekala}, {Somers}, {Grankin}, {Covey}, {Donati}, {Alencar}, {Hussain}, \&
  {Shappee}}]{gully17}
{Gully-Santiago}, M.~A., {Herczeg}, G.~J., {Czekala}, I., {et~al.} 2017, \apj,
  836, 200

\bibitem[{{Hatzes} \& {Rauer}(2015)}]{hatzes15}
{Hatzes}, A.~P., \& {Rauer}, H. 2015, \apjl, 810, L25

\bibitem[{{Herczeg} \& {Hillenbrand}(2014)}]{herczeg14}
{Herczeg}, G.~J., \& {Hillenbrand}, L.~A. 2014, \apj, 786, 97

\bibitem[{{Ingalls} {et~al.}(2016){Ingalls}, {Krick}, {Carey}, {Stauffer},
  {Lowrance}, {Grillmair}, {Buzasi}, {Deming}, {Diamond-Lowe}, {Evans},
  {Morello}, {Stevenson}, {Wong}, {Capak}, {Glaccum}, {Laine}, {Surace}, \&
  {Storrie-Lombardi}}]{ingalls16}
{Ingalls}, J.~G., {Krick}, J.~E., {Carey}, S.~J., {et~al.} 2016, \aj, 152, 44

\bibitem[{{Ireland} {et~al.}(2011){Ireland}, {Kraus}, {Martinache}, {Law}, \&
  {Hillenbrand}}]{ireland11}
{Ireland}, M.~J., {Kraus}, A., {Martinache}, F., {Law}, N., \& {Hillenbrand},
  L.~A. 2011, \apj, 726, 113

\bibitem[{{Janson} {et~al.}(2015){Janson}, {Quanz}, {Carson}, {Thalmann},
  {Lafreni{\`e}re}, \& {Amara}}]{janson15}
{Janson}, M., {Quanz}, S.~P., {Carson}, J.~C., {et~al.} 2015, \aap, 574, A120

\bibitem[{{Kratter} {et~al.}(2010){Kratter}, {Murray-Clay}, \&
  {Youdin}}]{kratter10}
{Kratter}, K.~M., {Murray-Clay}, R.~A., \& {Youdin}, A.~N. 2010, \apj, 710,
  1375

\bibitem[{{Kraus} {et~al.}(2015){Kraus}, {Andrews}, {Bowler}, {Herczeg},
  {Ireland}, {Liu}, {Metchev}, \& {Cruz}}]{kraus15}
{Kraus}, A.~L., {Andrews}, S.~M., {Bowler}, B.~P., {et~al.} 2015, \apjl, 798,
  L23

\bibitem[{{Kraus} {et~al.}(2017){Kraus}, {Herczeg}, {Rizzuto}, {Mann},
  {Slesnick}, {Carpenter}, {Hillenbrand}, \& {Mamajek}}]{kraus17}
{Kraus}, A.~L., {Herczeg}, G.~J., {Rizzuto}, A.~C., {et~al.} 2017, \apj, 838,
  150

\bibitem[{{Kraus} \& {Hillenbrand}(2009{\natexlab{a}})}]{kraus09b}
{Kraus}, A.~L., \& {Hillenbrand}, L.~A. 2009{\natexlab{a}}, \apj, 703, 1511

\bibitem[{{Kraus} \& {Hillenbrand}(2009{\natexlab{b}})}]{kraus09a}
---. 2009{\natexlab{b}}, \apj, 704, 531

\bibitem[{{Kraus} \& {Hillenbrand}(2012)}]{kraus12}
---. 2012, \apj, 757, 141

\bibitem[{{Kraus} {et~al.}(2014){Kraus}, {Ireland}, {Cieza}, {Hinkley},
  {Dupuy}, {Bowler}, \& {Liu}}]{kraus14}
{Kraus}, A.~L., {Ireland}, M.~J., {Cieza}, L.~A., {et~al.} 2014, \apj, 781, 20

\bibitem[{{Lafreni{\`e}re} {et~al.}(2008){Lafreni{\`e}re}, {Jayawardhana},
  {Brandeker}, {Ahmic}, \& {van Kerkwijk}}]{lafreniere08}
{Lafreni{\`e}re}, D., {Jayawardhana}, R., {Brandeker}, A., {Ahmic}, M., \& {van
  Kerkwijk}, M.~H. 2008, \apj, 683, 844

\bibitem[{{Lawrence} {et~al.}(2012){Lawrence}, {Warren}, {Almaini}, {Edge},
  {Hambly}, {Jameson}, {Lucas}, {Casali}, {Adamson}, {Dye}, {Emerson},
  {Foucaud}, {Hewett}, {Hirst}, {Hodgkin}, {Irwin}, {Lodieu}, {McMahon},
  {Simpson}, {Smail}, {Mortlock}, \& {Folger}}]{lawrence12}
{Lawrence}, A., {Warren}, S.~J., {Almaini}, O., {et~al.} 2012, VizieR Online
  Data Catalog, 2314

\bibitem[{{Low} \& {Lynden-Bell}(1976)}]{low76}
{Low}, C., \& {Lynden-Bell}, D. 1976, \mnras, 176, 367

\bibitem[{{Luhman}(2004)}]{luhman04}
{Luhman}, K.~L. 2004, \apj, 602, 816

\bibitem[{{Luhman} {et~al.}(2010){Luhman}, {Allen}, {Espaillat}, {Hartmann}, \&
  {Calvet}}]{luhman10}
{Luhman}, K.~L., {Allen}, P.~R., {Espaillat}, C., {Hartmann}, L., \& {Calvet},
  N. 2010, \apjs, 186, 111

\bibitem[{{Luhman} \& {Mamajek}(2012)}]{luhman12}
{Luhman}, K.~L., \& {Mamajek}, E.~E. 2012, \apj, 758, 31

\bibitem[{{Luhman} {et~al.}(2009){Luhman}, {Mamajek}, {Allen}, {Muench}, \&
  {Finkbeiner}}]{luhman09}
{Luhman}, K.~L., {Mamajek}, E.~E., {Allen}, P.~R., {Muench}, A.~A., \&
  {Finkbeiner}, D.~P. 2009, \apj, 691, 1265

\bibitem[{{Marengo} {et~al.}(2006){Marengo}, {Megeath}, {Fazio}, {Stapelfeldt},
  {Werner}, \& {Backman}}]{marengo06}
{Marengo}, M., {Megeath}, S.~T., {Fazio}, G.~G., {et~al.} 2006, \apj, 647, 1437

\bibitem[{{Pearce} {et~al.}(2019){Pearce}, {Kraus}, {Dupuy}, {Ireland},
  {Rizzuto}, {Bowler}, {Birchall}, \& {Wallace}}]{pearce19}
{Pearce}, L.~A., {Kraus}, A.~L., {Dupuy}, T.~J., {et~al.} 2019, \aj, 157, 71

\bibitem[{{Pecaut} {et~al.}(2012){Pecaut}, {Mamajek}, \& {Bubar}}]{pecaut12}
{Pecaut}, M.~J., {Mamajek}, E.~E., \& {Bubar}, E.~J. 2012, \apj, 746, 154

\bibitem[{{Pollack} {et~al.}(1996){Pollack}, {Hubickyj}, {Bodenheimer},
  {Lissauer}, {Podolak}, \& {Greenzweig}}]{pollack96}
{Pollack}, J.~B., {Hubickyj}, O., {Bodenheimer}, P., {et~al.} 1996, \icarus,
  124, 62

\bibitem[{{Preibisch} {et~al.}(2002){Preibisch}, {Brown}, {Bridges},
  {Guenther}, \& {Zinnecker}}]{preibisch02}
{Preibisch}, T., {Brown}, A.~G.~A., {Bridges}, T., {Guenther}, E., \&
  {Zinnecker}, H. 2002, \aj, 124, 404

\bibitem[{{Preibisch} {et~al.}(1998){Preibisch}, {Guenther}, {Zinnecker},
  {Sterzik}, {Frink}, \& {Roeser}}]{preibisch98}
{Preibisch}, T., {Guenther}, E., {Zinnecker}, H., {et~al.} 1998, \aap, 333, 619

\bibitem[{{Schneider} {et~al.}(2011){Schneider}, {Dedieu}, {Le Sidaner},
  {Savalle}, \& {Zolotukhin}}]{schneider11}
{Schneider}, J., {Dedieu}, C., {Le Sidaner}, P., {Savalle}, R., \&
  {Zolotukhin}, I. 2011, \aap, 532, A79

\bibitem[{{Scholz} {et~al.}(2012){Scholz}, {Stelzer}, {Costigan}, {Barrado},
  {Eisl{\"o}ffel}, {Lillo-Box}, {Riviere-Marichalar}, \& {Stoev}}]{scholz12}
{Scholz}, A., {Stelzer}, B., {Costigan}, G., {et~al.} 2012, \mnras, 419, 1271

\bibitem[{{Schwarz} {et~al.}(2016){Schwarz}, {Ginski}, {de Kok}, {Snellen},
  {Brogi}, \& {Birkby}}]{schwarz16}
{Schwarz}, H., {Ginski}, C., {de Kok}, R.~J., {et~al.} 2016, \aap, 593, A74

\bibitem[{{Silk}(1977)}]{silk77}
{Silk}, J. 1977, \apj, 214, 152

\bibitem[{{Slesnick} {et~al.}(2006){Slesnick}, {Carpenter}, {Hillenbrand}, \&
  {Mamajek}}]{slesnick06}
{Slesnick}, C.~L., {Carpenter}, J.~M., {Hillenbrand}, L.~A., \& {Mamajek},
  E.~E. 2006, \aj, 132, 2665

\bibitem[{{Stelzer} {et~al.}(2010){Stelzer}, {Scholz}, {Argiroffi}, \&
  {Micela}}]{stelzer10}
{Stelzer}, B., {Scholz}, A., {Argiroffi}, C., \& {Micela}, G. 2010, \mnras,
  408, 1095

\bibitem[{{Thompson} {et~al.}(2018){Thompson}, {Coughlin}, {Hoffman},
  {Mullally}, {Christiansen}, {Burke}, {Bryson}, {Batalha}, {Haas},
  {Catanzarite}, {Rowe}, {Barentsen}, {Caldwell}, {Clarke}, {Jenkins}, {Li},
  {Latham}, {Lissauer}, {Mathur}, {Morris}, {Seader}, {Smith}, {Klaus},
  {Twicken}, {Van Cleve}, {Wohler}, {Akeson}, {Ciardi}, {Cochran}, {Henze},
  {Howell}, {Huber}, {Pr{\v s}a}, {Ram{\'{\i}}rez}, {Morton}, {Barclay},
  {Campbell}, {Chaplin}, {Charbonneau}, {Christensen-Dalsgaard}, {Dotson},
  {Doyle}, {Dunham}, {Dupree}, {Ford}, {Geary}, {Girouard}, {Isaacson},
  {Kjeldsen}, {Quintana}, {Ragozzine}, {Shabram}, {Shporer}, {Silva Aguirre},
  {Steffen}, {Still}, {Tenenbaum}, {Welsh}, {Wolfgang}, {Zamudio}, {Koch}, \&
  {Borucki}}]{thompson18}
{Thompson}, S.~E., {Coughlin}, J.~L., {Hoffman}, K., {et~al.} 2018, \apjs, 235,
  38

\bibitem[{{Tomida} {et~al.}(2013){Tomida}, {Tomisaka}, {Matsumoto}, {Hori},
  {Okuzumi}, {Machida}, \& {Saigo}}]{tomida13}
{Tomida}, K., {Tomisaka}, K., {Matsumoto}, T., {et~al.} 2013, \apj, 763, 6

\bibitem[{{Torres} {et~al.}(2009){Torres}, {Loinard}, {Mioduszewski}, \&
  {Rodr{\'{\i}}guez}}]{torres09}
{Torres}, R.~M., {Loinard}, L., {Mioduszewski}, A.~J., \& {Rodr{\'{\i}}guez},
  L.~F. 2009, \apj, 698, 242

\bibitem[{{Voirin} {et~al.}(2018){Voirin}, {Manara}, \& {Prusti}}]{voirin18}
{Voirin}, J., {Manara}, C.~F., \& {Prusti}, T. 2018, \aap, 610, A64

\bibitem[{{Vorobyov}(2013)}]{vorobyov13}
{Vorobyov}, E.~I. 2013, \aap, 552, A129

\bibitem[{{Wagner} {et~al.}(2019){Wagner}, {Apai}, \& {Kratter}}]{wagner19}
{Wagner}, K., {Apai}, D., \& {Kratter}, K.~M. 2019, \apj, 877, 46

\bibitem[{{Werner} {et~al.}(2004){Werner}, {Roellig}, {Low}, {Rieke}, {Rieke},
  {Hoffmann}, {Young}, {Houck}, {Brandl}, {Fazio}, {Hora}, {Gehrz}, {Helou},
  {Soifer}, {Stauffer}, {Keene}, {Eisenhardt}, {Gallagher}, {Gautier}, {Irace},
  {Lawrence}, {Simmons}, {Van Cleve}, {Jura}, {Wright}, \&
  {Cruikshank}}]{werner04}
{Werner}, M.~W., {Roellig}, T.~L., {Low}, F.~J., {et~al.} 2004, \apjs, 154, 1

\bibitem[{{Wichmann} {et~al.}(1996){Wichmann}, {Krautter}, {Schmitt},
  {Neuhaeuser}, {Alcala}, {Zinnecker}, {Wagner}, {Mundt}, \&
  {Sterzik}}]{wichmann96}
{Wichmann}, R., {Krautter}, J., {Schmitt}, J.~H.~M.~M., {et~al.} 1996, \aap,
  312, 439

\bibitem[{{Wright} {et~al.}(2011){Wright}, {Fakhouri}, {Marcy}, {Han}, {Feng},
  {Johnson}, {Howard}, {Fischer}, {Valenti}, {Anderson}, \&
  {Piskunov}}]{wright11}
{Wright}, J.~T., {Fakhouri}, O., {Marcy}, G.~W., {et~al.} 2011, \pasp, 123, 412

\bibitem[{{Wu} {et~al.}(2017){Wu}, {Close}, {Eisner}, \& {Sheehan}}]{wu17b}
{Wu}, Y.-L., {Close}, L.~M., {Eisner}, J.~A., \& {Sheehan}, P.~D. 2017, \aj,
  154, 234

\bibitem[{{Wu} \& {Sheehan}(2017)}]{wu17a}
{Wu}, Y.-L., \& {Sheehan}, P.~D. 2017, \apjl, 846, L26

\bibitem[{Zhou {et~al.}(2014)Zhou, Herczeg, Kraus, Metchev, \& Cruz}]{zhou14}
Zhou, Y., Herczeg, G.~J., Kraus, A.~L., Metchev, S., \& Cruz, K.~L. 2014, The
  Astrophysical Journal Letters, 783, L17

\end{thebibliography}
\end{document}